\documentclass[aps, prd, amsmath, floats, floatfix, twocolumn, superscriptaddress, nofootinbib, showpacs]{revtex4}
 
\usepackage{amssymb}
\usepackage{amsmath}
\usepackage{verbatim}
\usepackage{mathrsfs}
\usepackage{amsfonts}
\usepackage{latexsym}
\usepackage{epsfig}
\usepackage{epstopdf}
\usepackage[usenames, dvipsnames]{color}
\usepackage{hyperref}
\hypersetup{
    colorlinks=true,
    linkcolor=blue,
    citecolor=blue,
    urlcolor=red,
    pdftitle={Fermi bubbles in SFDM},
    }

\definecolor{pink}{rgb}{0.9,0,0.6} 

\newcommand{\beq}{\begin{equation}}
\newcommand{\eeq}{\end{equation}}
\newcommand{\bea}{\begin{eqnarray}}
\newcommand{\eea}{\end{eqnarray}}
\newcommand{\beqn}{\begin{equation*}}
\newcommand{\eeqn}{\end{equation*}}
\newcommand{\bean}{\begin{eqnarray*}}
\newcommand{\eean}{\end{eqnarray*}}

\long\def\symbolfootnote[#1]#2{\begingroup%
\def\thefootnote{\fnsymbol{footnote}}\footnote[#1]{#2}\endgroup}

\begin{document}

\title[The Quantum character of SFDM]{The quantum character of the Scalar Field Dark Matter}


\author{Tonatiuh Matos}\email[0000-0002-0570-7246 ]{tonatiuh.matos@cinvestav.mx}
\affiliation{Departamento de F\'isica, Centro de Investigaci\'on y de Estudios Avanzados del IPN, A.P. 14-740, 07000 CDMX.,  M\'exico.}




\begin{abstract}
The Scalar Field Dark Matter (SFDM) model, also called Fuzzy, Wave, Bose-Einstein, Ultra-light Dark Matter, has received a lot of attention because it has been able to provide simpler and more natural explanations for various features of galaxies, such as the number of satellite galaxies and the cusp-core problem. We recently showed that this model is able to explain the vast polar orbits of satellite galaxies around their host, the so-called VPO, and to explain the X-ray and gamma-ray emissions in the vacuum regions of our galaxy, that is, the Fermi Bubbles. In all these phenomena the quantum character of SFDM has been crucial. In this work we study the quantum effects of SFDM at the cosmological level, to see these effects not only at the galactic scale, but also at the cosmological scale. Using a convenient ansatz, we were able to integrate the perturbed equations to show that the shape of the SFDM halos resembling atoms is a generic result. The main conclusion of this work is that quantum mechanics, the successful microworld theory, could also explain the dark side of the cosmos.
\end{abstract}


\maketitle




\section{Introduction}
\label{sec:introduction}

 Dark Matter (DM) is considered one of the most important scientific challenges to be solved in this century. So far more than 95\% of the matter in the universe is unknown, of which more than a quarter is DM, some kind of force that forms the structure on large scales, from superclusters of galaxies to dwarf galaxies in the universe. Since its discovery in the early 1930s, the question of what the matter in the universe is made of has been a challenge that remains one of the most important unsolved mysteries in science.
 
 The cold dark matter (CDM) model suffers from a number of challenges in terms of its predictions, especially on galactic scales, and this has motivated the search for alternatives more consistent with the observations (see for example \cite{Suarez:2013iw}, \cite{Oks:2021hef}).

 One of them, which we will address here, is the Scalar Field Dark Matter (SFDM) model. In 1998 we proposed that the DM is a scalar field, that is, a particle with spin 0, satisfying the Klein-Gordon equations that drives the dynamics of the universe. In this work \cite{Matos:1998vk} we show that this hypothesis could explain the observed rotation curves of the stars and gas around the galaxies. After that, the SFDM idea has been rediscovered many times such as Fuzzy (\cite{Hu:2000ke}), Bose-Einstein (\cite{Boehmer:2007um}; \cite{Rindler-Daller:2009qyu}), Wave DM (\cite{Bray:2010fc}; \cite{Schive:2014dra}), etc. This idea started to be a fad and one of the favorite candidates to explain DM (\cite{Hui:2016ltb}).
Shortly after, Matos and Ureña-L\'opez studied the same hypothesis for the first time, but now from a cosmological point of view in \cite{Matos:2000ss}. The results were spectacular, finding for the first time that all cosmological observations up to that point were explained within the error bars by the SFDM. Some of the main results of this work are:

1.- Using the mass of the scalar field as a free parameter, in this work we showed that the scalar field has a natural cutoff of the mass power spectrum, which implies that the theoretical number of satellite galaxies is of the order of magnitude of the observed ones. This result was corroborated many years later using numerical simulations (\cite{Schive:2014dra}).

2.- The Mass Power Spectrum and the Angular Power Spectrum (CMB spectrum) agree with the theoretical results of the model, something that was corroborated many years later (\cite{Hlozek:2014lca}).
That means that the Scalar Field Dark Matter model can explain the galaxies, the number of satellite galaxies in the big ones, and all the observations made up to that point on cosmological scales.

Years later, we realize for the first time that there were two problems with the model that we had to deal with. The first was discovered in \cite{Guzman:2003kt}. Here it was found that galaxies could be unstable according to this model. And second, that supermassive black holes at the center of galaxies could swallow the entire scalar field. This last problem was addressed for the first time in \cite{Urena-Lopez:2002nup} and latter in \cite{Avilez:2017jql} and \cite{Padilla:2020sjy}, where they found that supermassive black holes can coexist with the scalar field halo. This result was corroborated several times later (\cite{Barranco:2011eyw}).
The first problem was addressed for the first time in the literature considering the quantum characteristics of the scalar field using the excited states of the system. This idea started a new paradigm in the literature called l-boson stars (\cite{Alcubierre:2018ahf}). The problem was attacked in two ways. The first way was to consider the scalar field system to have several wavefunctions as states. In \cite{Urena-Lopez:2010zva} the properties of gravitationally bound multistate configurations made of spin-zero bosons in the Newtonian regime are studied in detail. They find that the system is stable if it contains several states at the same time. This result was corroborated in \cite{Guzman:2019gqc} and more recently in \cite{Guzman:2022vxl}. This result is very important for this work. This same problem was addressed with quantum field theory at finite temperature in \cite{Matos:2011pd}; \cite{Robles:2012kt} and \cite{Robles:2013ioa}.


{SFDM could have various origins, for example, in \cite{Hui:2016ltb} the authors propose that SFDM is derived from superstring theory. Here the scalar field is real and has no charge. It is also possible to propose that we add the SFDM Lagrangian to the Standard Model (SM) of particle, or another possibility may be a complex scalar field $\Phi$ (\cite{Rindler-Daller:2013zxa}). It can also be proposed that the complex scalar field is in a thermal bath at temperature $T$ endowed with an SFDM potential given by (see for example \cite{Robles:2012kt} for a real scalar field in a thermal bath)
\begin{equation}\label{eq:V}
    V =-m_\Phi^2\Phi\Phi^*+\frac{\lambda}{2}(\Phi\Phi^*)^2+\frac{\lambda}{4}\Phi\Phi^*T^2+\frac{\pi^2}{90} T ^4.
\end{equation}
In this case the scalar field is not charged and could have an interaction term with the SM with a super small interaction constant or the interaction constant being zero.
Another possibility is that the SFDM is charged and does not interact with any other SM components, but its corresponding dark photon component does interact with the SM photon. In this work this is the version that we will adopt, since this version can explain the anisotropic distribution of satellite galaxies and the super-energetic emissions in empty regions of the galaxy. Let us clarify these points. All dark matter models predict that the satellite galaxies should move in the host galaxy in homogeneous orbits, all of them uniformly distributed. However, until now, current telescopes have seen the satellite galaxies of three galaxies, the Milky Way  (\cite{Pawlowski:2013cae}; \cite{Pawlowski:2019bar}), Andromeda (\cite{Ibata:2013rh}; \cite{Conn:2013iu}) and Sagittarius A  (\cite{Muller:2018hks}; \cite{Muller:2020njt}) and in all of them the satellites are not evenly distributed. This phenomenon is called Vast Polar Orbits (VPO) in galaxies.
The other challenge is to explain the radiation observed in the almost empty intergalactic regions, where NASA's Fermi satellite observes continuous emanations of X-rays, gamma rays and even more energetic than that, without any reasonable explanation, the so-called Fermi Bubbles (FB) (\cite{Su:2010qj}). 

Then the idea is the following.
At the beginning of the universe, the SFDM was in thermal equilibrium with the other components of the SM, but decoupled from the SM very early in the universe's history. Observations show that the self-interaction $\lambda$ to meet the constraints of nucleosynthesis must be very small, but different from zero (\cite{Li:2013nal}). At the same time, the charge of the SFDM must be very small to comply with the FB constraints (\cite{Solis-Lopez:2019lvz}). Like the rest of the components, SFDM cools down and makes a phase transition at $T_c=2m_\Phi/\sqrt{\lambda}$, very early in the history of the universe due to its tiny interaction with itself $\lambda$ (for this part of the idea you can neglect the small SFDM charge). Now we follow the standard model of cosmology. After inflation, quantum fluctuations become classical and collapse the SFDM to form the cosmic structure of the universe (\cite{Matos:2000ss}; \cite{Urena-Lopez:2010zva}; \cite{Hlozek:2014lca}). Due to the expansion, the volume of the fluctuations increases causing the scalar field to cool to form Bose-Einstein condensates which at the same time form halos of galaxies, where most of the SFDM particles go to the ground state. But due to gravity, the fluctuations begin to collapse again the scalar field, causing it a turn around. So the volume of the fluctuations decreases and the temperature of the halo increases, causing some of the SFDM particles to enter excited states.
In \cite{Guzman:2019gqc} it was shown that if most of the particles remain in the ground state, but with a part of them in at least one excited state, the system stabilizes and remains stable for at least the age of the universe, provided the mass of the SFDM particle is small enough. They also showed that if this does not happen the system becomes unstable and disappears. 

The mass of SFDM is ultralight, which means that any perturbation in its temperature is comparable to this mass in energy. In \cite{Robles:2012kt}, it was shown that for a scalar field in a thermal bath, the characteristic polynomial is given by (equation (15) in this reference) 
\begin{equation}
    \omega^2=k^2c^2+m_\Phi^2c^2\left(1-\frac{T^2}{T_c^2}\right),
\end{equation}
where $m_\Phi$ is the mass of the scalar field, $T$ is its temperature and $T_c=2m_\Phi/\sqrt{\lambda}$ is its critical phase transition temperature.
We can assume that the mass of the scalar field is $m_\Phi\sim 10^{-21}$eV, which agrees with the constrictions of Ly-alpha and of the satellite galaxies of the Milky Way (\cite{DES:2020fxi}). 
But because the scalar field in a galaxy halo heats up, the effective mass decreases as $m^2=m^2_\Phi(1-T^2/T_c^2)$. Therefore, galaxies with different sizes show different effective masses. In the case of large galaxies, such as the Milky Way, this effective mass is of the order of $m\sim10^{-24}$eV. This implies that the halo of a galaxy is still a Bose-Einstein condensate, so it continues to behave like DM, but now with some particles in excited states.}

The idea, then, is to give the SFDM a small charge and take into account the excited states of the halo of the galaxy, treating the halo of the galaxy as an atom, to explain the new observations that have not been explained so far, such as the VPO (\cite{Solis-Lopez:2019lvz}) and the FB (\cite{Matos:2022ruq}). For doing so, we consider the internal symmetry of the SFDM to be the group $U(1)$. The corresponding Lagrangian then contains a new charge $q$, which we assume must be dark. {This guarantees that the SFDM particles do not interact with the rest of the SM components, as an observed feature of DM, avoiding any conflict with the SM, only the SFDM photon interacts with the SM photon, an interaction that manifests as FB.} Because of this, the SFDM can interact with the normal SM electromagnetic field, but not (directly) with other particles. In this case, \cite{Matos:2022ruq} showed that due to SFDM colliding with photons from starlight or the CMB, these photons acquire enough energy to be seen as X-rays, gamma rays, or even more energetic radiation. This is enough to give an alternative explanation of what we see as FB. 

Thus, we start with the Lagrangian
\begin{eqnarray}\label{eq:Lagrangian0}
\mathcal{L}&=&(\nabla_\mu\Phi+iqB_\mu\Phi)(\nabla^\mu\Phi^*-iqB^\mu\Phi^*)-m_\Phi^2\Phi\Phi^*\nonumber\\
&-&\frac{1}{4}B_{\mu\nu}B^{\mu\nu}-\frac{1}{4}B^\prime_{\mu\nu}B^{\prime\mu\nu}-\frac{\delta}{2}B_{\mu\nu}B^{\prime\mu\nu},
\end{eqnarray}
where $B_\mu$ is the dark gauge field
of the SFDM, with fundamental charge $q$ and Faraday tensor $B_{\mu\nu}=B_{\mu;\nu}-B_{\nu;\mu}$,  whereas $B^\prime_{\mu\nu}=B^\prime_{\mu;\nu}-B^\prime_{\nu;\mu}$ 
is the electromagnetic gauge field of the SM 
and $\delta$  a kinetic coupling constant between these two fields. By construction, there is not ad hoc coupling between the scalar field $\Phi$ or $B_\mu$ and the rest of the field components of the SM.
The scalar field is the dark matter of the universe and has the dominant part of the gravitational field, the electromagnetic fields are small and their contribution to the gravity of the system is negligible.
The scalar field satisfies the Klein-Gordon (KG) equation, however, in a galaxy it suffices to work with the non-relativistic limit of the KG equations which reduces to the Schr\"odinger equation. Therefore we can interpret the SFDM halos of galaxies as atoms. In the same limit, Einstein's equations reduce to Poisson's equation. Thus, to analyze the halo of a galaxy, the Einstein-KG field equations reduce to the Schrödinger-Poisson (SP) system. Quantum mechanics predicts the existence of atoms, and atoms have various quantum states, called $s$, $p$, $d$, etc. The $s$ states are spherical but the $p$ states are bubbles at the north and south poles of the atom and so on.
Numerical evolutions of this system have been performed in \cite{Guzman:2019gqc} and the result is that the scalar field density profile resembles that of the Legendre function $P_k^j(\cos(\theta))$ of the hydrogen atom, with quantum numbers $k$ and $j$, which implies that the SFDM profile behaves very similarly to a hydrogen atom. This reminds us very well of the shape of the FB in the galaxy and can well explain the anisotropic distribution of the satellite galaxies, the VPO.

Thus, SFDM has a very clear quantum behavior.
If we take into account the multistate characteristic of the Scalar Field, we can explain the behavior of the VPO in a very simple and natural way. This was shown in a recent paper (\cite{Solis-Lopez:2019lvz}). Furthermore, with this same structure, considering that the scalar field is endowed with a minuscule charge that behaves like a dark photon, the FB can be explained, again, in a very simple and natural way. This proposal was given in \cite{Matos:2022ruq}. The SFDM is the only model so far that can give a simple, natural and reasonable explanation of these two phenomena.

The main objective of this work is to study the quantum character of SFDM at cosmological scales. We do not intend to analyze the evolution of the fluctuation with the scalar field, this has already been done in \cite{Matos:2000ss}; \cite{Urena-Lopez:2015gur}; \cite{Hlozek:2014lca}; \cite{Medellin-Gonzalez:2020eww} and \cite{Foidl:2022bpn}. Instead, we want to explicitly show the quantum behavior of the scalar field during the evolution of the universe. In order to do that the paper is organized as follows. 
In section \ref{sec:2Fieleq} we write the basic field equations for a charged complex scalar field and perform a polar decomposition of the scalar field function $\Phi$ into its norm and its phase. This separation allows us to rewrite the KG equation as a continuity equation and another one that is the Bernoulli equation. This is called the hydrodynamic representation of the KG equation. This form of the KG equations allows us to point out that the only difference between a classical hydrodynamic system and the KG equation is a term called the quantum potential. We can interpret this result as if this term is zero, the system is classical. Therefore we can say that the quantum character of the KG equation is contained in this term. In section \ref{sec:2Fieleq} we separate the scalar field function $\Phi$ into its background part and a linear perturbation and write the field equations in terms of its norm and phase, as in section \ref{sec:2Fieleq}. We propose here the convenient ansatz that allows us to solve the perturbed field equations in some cases, in terms of the scale factor and the background scalar field norm. These solutions are a toy model, but it is very important that they show us that quantum character of the scalar field perturbation at the end of its collapse. With this result we see that the SFDM halos of galaxies are real atoms. 
In section \ref{sec:4medulung} we perform a Madelung transformation, which essentially separates the phase of the scalar field into a function and an evolution parameter. The gradient of this new function can be interpreted as the velocity of the scalar field particles as if they were a fluid. In section \ref{sec:5Fourier} we perform a Fourier transformation of the field equations to transform them into a dynamical system and be able to solve them numerically. Before solving these equations numerically, in the sections \ref{sec:6Bigual0} and \ref{sec:7Bno0} we deal with a toy model where we first ignore the electromagnetic field and, using the convenient ansatz, we solve the perturbed field equations. The main result here is that without and with electromagnetic field we show that the final collapse of the SFDM is a halo in the form of an atom, that as usual, its ground state is spherically symmetric, and its first exited state is like the states $p$ of an atom, contains bubbles that can explain the VPO and FB in galaxies. 

This situation holds for realistic collapses using numerical simulations, but these analytical solutions tell us that this result is generic. In section \ref{sec:8General} we show the numerical simulation of the dynamical system, not with the aim of re-analyzing the evolution of the fluctuations, but with the aim of showing the quantum character of these fluctuations on cosmological scales. We point out that this feature is the main difference between the SFDM paradigm and other models. This feature is just the main point that allows us to explain various DM observations especially at galactic scales. Finally in the section \ref{sec:9conclusions} we give some conclusions. In Appendix A we explain how to write the perturbed quantities plotted in the figures using unitless variables.


\section{The Field equations}\label{sec:2Fieleq}

To write the field equations derived from the Lagrangian (\ref{eq:Lagrangian0}), we will consider
a complex scalar field, $\Phi=\Phi(x^i,t)$, $i=1,2,3$, where the wave function $\Phi$ is assumed to describe a charged particle of spin zero with point charge $q$ of mass at rest $m_\Phi$ that is coupled to an external electromagnetic field that is described by the potential of the form (\ref{eq:V}). For the evolution of the scalar field on cosmological scales we can neglect the dark photon interaction, but not for the final collapsed SFDM halo, where the dark photon is important. Therefore, on cosmological scales the Lagrangian (\ref{eq:Lagrangian0}) reduces to
\begin{eqnarray}\label{eq:Lagrangian}
{\cal L}&=&(\nabla_\alpha\,\Phi+iqB_\alpha\Phi)\,(\nabla^\alpha\,\Phi^*-iqB^\alpha\Phi^*)\nonumber\\&-& V(|\Phi|)-\frac{1}{4}B^{\mu\nu}B_{\mu\nu}, 
\end{eqnarray}
where $B_{\mu\nu}=\nabla_\nu B_{\mu}-\nabla_\mu B_{\nu}$. After symmetry breaking, we can assume that the scalar field potential is given only by
\begin{equation}\label{eq:Vpotential}
V(\Phi\Phi^*)=m_\Phi^2\Phi\Phi^*+\frac{\lambda}{2}(\Phi\Phi^*)^2,
\end{equation}
where now the scalar field oscillates near the minimum of the potential.
$\lambda$ is a self interaction parameter with units of inverse-squared distance. The constant $m_\Phi=\frac{1}{\lambda_\Phi}$ is related
to the wavelength of the scalar field $\lambda_\Phi$, which
is also related to the mass of the scalar field $M_\Phi$ {given in grams} by $m_\Phi=\frac{M_\Phi\,c}{\hbar}$,
where $c$ is the speed of light and $\hbar$ is Plank's constant.

The corresponding stress energy tensor of the scalar field is then
\begin{eqnarray}\label{eq:Tmunu}
{T^\Phi}_{\mu\,\nu}&=&\frac{c^4}{16\,\pi\,G}\,\left[(\nabla_\mu\Phi+iqB_\mu\Phi)\,(\nabla_\nu{\Phi^*}-iqB_\nu\Phi^*)\right.\nonumber\\
&+& \left.(\nabla_\mu{\Phi^*}-iqB_\mu\Phi^*)\,(\nabla_\nu\Phi+iqB_\nu\Phi)\right.\nonumber\\
&-&\left. g_{\mu\,\nu}\left(
g^{\alpha\,\beta}\,(\nabla_\alpha\Phi+iqB_\alpha\Phi)\,(\nabla_\beta{\Phi^*}-iqB_\beta\Phi^*)\right.\right. \nonumber\\
&+&\left. \left.m_\Phi^2\,\Phi\,\Phi^* + \frac{\lambda}{2}\,\left(\Phi\,\Phi^*\right)^2\right)\right],
\end{eqnarray}
The scalar field has units such that $\kappa\Phi$ is unitless, where $\kappa^2=8\pi G/c^4$ is Einstein's constant.

We first write the field equations derived from (\ref{eq:Lagrangian}) (see \cite{Matos:2016ryp}). To do this, we are going to
define the d'Alembert electromagnetic operator as
\begin{equation}
      \Box_\mathrm{E}=\left(\nabla^{\mu}+ i
q B^{\mu} \right) \left( \nabla_{\mu} + i q B_{\mu} \right),
\end{equation}
where $B_{\mu}$ is the vector field of the $U(1)$ gauge corresponding to Maxwell 4-potential, such
that the Klein-Gordon (KG) field equations are given by 
\begin{eqnarray}
	&&\Box_\mathrm{E} \Phi - \frac{\mathrm{d}V}{\mathrm{d}\Phi^*} = 0 ,
    \label{eq:KG} \\
 &&\nabla_\nu B^{\nu\mu} =J^{E\,\mu} ,
\label{eq:Maxwell}
\end{eqnarray}
for the complex scalar field $\Phi(\mathbf{x},t)$ and its complex conjugate
$\Phi^*(\mathbf{x},t)$. The conserved 4-current is defined as
\begin{eqnarray}
J^E_\mu \equiv i\frac{q}{2 m_\Phi^2} \left[\Phi
\left(\nabla_\mu - i q B_\mu \right)
      \Phi^* - \Phi^* \left( \nabla_\mu + i q B_\mu \right) \Phi \right] .
\label{eq:Jmu}
\end{eqnarray}
This describes a system of bosonic excitations that condense into a single macroscopic state.

In what follows we carry out the Madelung transformation defined as
\begin{eqnarray}
\Phi(\mathbf{x},t) = \sqrt{n} \exp(i\theta) = \sqrt{n}\exp[i(S - \omega_0 t)],
\label{eq:definition}
\end{eqnarray}
where the complex function $\Phi$ decomposes as any complex function in its norm $n(\mathbf{x},t)$ and a phase
$\theta(\mathbf{x},t)$. In this way, the Einstein-KG equation is divided into its imaginary and real parts, respectively,
\begin{equation}
	\nabla_{\mu} \sqrt{n} ( 2 \nabla^{\mu} \theta  + q B^{\mu} )
    + q \nabla_{\mu} ( B^{\mu} \sqrt{n} ) + \sqrt{n} \, \Box \theta = 0 ,
\label{GP:imag}
\end{equation}
\begin{equation}
	\Box \sqrt{n} - \sqrt{n} \left[ \nabla_{\mu} \theta ( \nabla^{\mu}
\theta  + 2 q B^{\mu} ) + q^2 B^2 +m_\Phi^2 + \lambda n \right] = 0 ,
	\label{GP:real}
\end{equation}
where $B^2 = B^\mu B_\mu$. Using these results, the
current (\ref{eq:Jmu}) is transformed into
\begin{equation}
 	J^E{}_{\mu}= \frac{n q}{m_\Phi^2}\left( \nabla_{\mu}
\theta + q B_{\mu} \right).
 \label{eq:JE}
 \end{equation}
  
 In terms of 
$J^E{}_{\mu}$,
Eqs.~\eqref{GP:imag} and \eqref{GP:real} read
\begin{eqnarray}
	\nabla^\mu J^E{}_\mu &=& 0 ,
\label{eq:nablaJE}\\
	J^E{}_{\mu}J^E{}^\mu  + \frac{n^2q^2}{{ m_\Phi}^4}
\left(  m_\Phi^2 + \lambda n
	- \frac{\Box \sqrt{n}}{\sqrt{n}} \right) &=& 0 .
\label{eq:realJE}
\end{eqnarray}
So, interpreting the KG equation
via the Madelung transformation, it splits into the continuum \eqref{eq:nablaJE} and the
Quantum Hamilton-Jacobi equations \eqref{eq:realJE} above. The quantum version
of the Hamilton-Jacobi equation differs from the classical one only by the last term in
the left side of Eq. \eqref{eq:realJE}, (see \cite{Matos:2016ryp})
\begin{equation}\label{eq:UQ}
  U^Q=-\frac{1}{2m_\Phi^2}\frac{\Box \sqrt{n}}{\sqrt{n}},
\end{equation}
which corresponds to
de Broglie's relativistic quantum
potential (\cite{Bohm:1951xx}). This is the fundamental difference between a hydrodynamic system and a quantum system. In this work we will interpret this term as the contribution of the quantum mechanical part of the field equations, if it is zero, the system is a classical system, if not, this term will give us the contribution of the quantum field.

On the other hand, it is useful to make the following analogy. In a flat spacetime of a gaseous boson, the 4-electromagnetic moment corresponds to the sum of the individual mechanical moments, namely $
J^E{}_{\mu} = ( q/ m_\Phi) n v_\mu$. Therefore, we introduce the speed
\begin{equation}
	 m_\Phi v_\mu \equiv \nabla_\mu S +  q B_\mu .
\label{eq:vel}
\end{equation}

In terms of $v_\mu$, the continuity and quantum
Hamilton-Jacobi equations \eqref{eq:nablaJE} and \eqref{eq:realJE} become
\begin{eqnarray}
	\nabla^\mu (nv_\mu) - \frac{\omega_0}{ m_\Phi}  \left( \nabla^0 n + n \, \Box \,
    t \right) = 0 ,\quad
\label{eq:n1} \\
	v_\mu v^\mu - \frac{2 \omega_0}{m_\Phi} v^0 - \frac{\omega_0^2}{ m_\Phi^2 N^2} +
   1 + \frac{\lambda}{ m_\Phi^2} n - \frac{1}{ m_\Phi^2}\frac{\Box\sqrt{n}} {\sqrt{n}} = 0
.\quad
\label{eq:n2}
\end{eqnarray}
The equation \eqref{eq:n1} is the generalized continuity equation of hydrodynamics that governs the density evolution of the boson gas, while the equation \eqref{eq:n2} is the generalized Bernoulli equation that governs the evolution of its velocity.

\section{The perturbed Field Equations}\label{sec:3Pertubed}

In the present work we follow the standard idea that the universe is homogeneous and isotropic and the structure of the universe is due to fluctuations that grow throughout the history of the universe. We can separate these fluctuations from the homogeneous and isotropic background by separating the metric into
\begin{equation}
      g_{\mu\nu}=\eta_{\mu\nu}+h_{\mu\nu},
\end{equation}
where the perturbed metric $h_{\mu\nu}<<1$. In that case, the Einstein tensor $G^\mu_\nu$ and the energy moment tensor $T^\mu_\nu$ can be separated as
\begin{equation}
G^{\mu}_{\nu}=G^{0\mu}_{\,\,\nu} +\delta {G}{^\mu_\nu},\,\,\,\,\,\, {T}{^\mu_\nu}=T^{0\mu}_{\,\,\nu} +\delta{T}{^\mu_\nu},
\end{equation}
being $G^{0\mu}_{\,\,\nu}$ the Einstein tensor for the background and $\delta {G}{^\mu_\nu}$ for the perturbation, analogously for the energy momentum tensor. 

For the homogeneous and isotropic background we use the Friedman - Lem\^aitre line element in its conformal time representation
\begin{equation}
ds^2=a^2(\eta)\left(-c^2d\eta^2 + \frac{dr^2}{\sqrt{1-kr^2}} +d\,\Omega^2\right),
\label{eq:ele_F}
\end{equation}
where $d\Omega^2=d\,\theta^2 + \sin^2\theta d\varphi$, and
$a(\eta)$ is the scale factor which is just a function of $\eta$. The background metric can be solved independently of the rest of the equation. The equation for the perturbation is
\begin{equation} \label{eq:deltaG=deltaT}
\delta G{^\mu_\nu}=\kappa^2\delta T{^\mu_\nu},
\end{equation}

Here we will consider the Newtonian gauge metric only with scalar perturbations. In that case, the perturbed metric can be written as
\begin{equation}\label{gauge1234}
\text{d}s^2= a^2(\eta)\,\left[-(1+2\psi)\,\text{d}\eta^2+ (1-2\phi)\eta_ {ij}\,\text{d}x^{i}\,\text{d}x^{j}\right].
\end{equation}
while the scalar field can be separated into the background part and a perturbed one as
\begin{equation}\label{eq:PhiPerturbation}
\Phi= \Phi_0\,(\eta)+ \delta \Phi\,(x^i,\eta),
\end{equation}
where $\Phi_0\,(\eta)$ is the background scalar field and $\delta\Phi<<1$ is a linear perturbation.
Using this metric, the Einstein's equations reduce to
\begin{eqnarray}
\frac{\dot a^2}{a^2}-\frac{2\ddot a}{a}&=&\kappa^2\left(\dot\Phi_0\dot\Phi_0^\ast-Va^ 2\right),\nonumber\\
\frac{3\dot a^2}{a^2}&=&\kappa^2\left(\dot\Phi_0\dot\Phi_0^\ast+Va^2\right),\label{eq:G00L}
\end{eqnarray}
for the background. Here dot means derivative with respect to $\eta$, $i.e.$, ${}^{\cdot} = d/d\eta$. It's easy to see that
\begin{equation}\label{eq:G00LP}
    \left(\frac{3\dot a}{a}\right)^\cdot=\kappa^2\left(-2\dot\Phi_0\dot\Phi_0^\ast+Va^2\right).
\end{equation}

The density $\rho_{\Phi_0}$ and the pressure $p_{\Phi_0}$ of the scalar field are defined as
\begin{eqnarray}
\rho_{\Phi_0}&=&\dot\Phi_0\dot\Phi_0^\ast/a^2+V,\label{eq:rhoPhi0}\\
      p_{\Phi_0}&=&\dot\Phi_0\dot\Phi_0^\ast/a^2-V\label{eq:pphi0}.
\end{eqnarray}

From the conservation of the energy moment tensor (\ref{eq:Tmunu}) we obtain the KG equation for $\Phi$
\begin{equation}
      \Box\Phi-\frac{dV}{d\Phi^\ast}=0,
\end{equation}
and the corresponding equation for $\Phi^\ast$. This equation for the background is the {time-varying, minimally coupled KG equation}
\begin{equation}\label{eq:KGhomogenea}
\ddot{\Phi}_0+2H\dot{\Phi}_0+V_{,\Phi_0^{\ast}}a^2=0,
\end{equation}
Using equation (\ref{eq:G00LP}) we can obtain the derivative of the density $\rho_{\Phi_0}$, we obtain
\begin{equation}\label{eq:rhoP}
    \dot\rho_{\Phi_0}=-6H\frac{\dot\Phi_0\dot\Phi_0^{\ast}}{a^2}.
\end{equation}

In what follows we derive the field equations for the perturbations of the scalar, metric and electromagnetic fields, we get
\begin{eqnarray}\label{eq:deltaG}
\delta{G}{^0_0}&=&6H\left(\dot{\phi}+H\psi\right)-\frac{2}{a^2}\nabla^2\phi,\label{eq:deltaG00}\\
\delta{G}{^0_j}&=&-2\left(\dot{\phi}+H\psi\right)_{,j}\label{eq:deltaG0j},\\
\delta{G}{^i_j}&=&2\left[\ddot{\phi}+H\left(2\dot{\phi}+\dot{\psi}\right)+\left(2\dot{H}+H^2 \right)\psi\right]\delta^i_j\nonumber\\
&-&\frac{2}{a^2}\left(\phi-\psi\right)^{,i}_{,j},\label{eq:deltaGij}
\end{eqnarray}
where, for simplicity, we have defined the parameter $H$ as $H=\dot a/a$, {but this is not the Hubble parameter $\mathcal{H}$}. Note that $H=1/a\,da/d\eta=da/dt$, where $t$ is the cosmological time $dt=ad\eta$, thus the Hubble parameter is given by $\mathcal{H}=1/a\,da/dt=H/a$.

In terms of the Newtonian metric, the perturbed scalar field energy momentum tensor (\ref{eq:Tmunu}) reduces to
\begin{eqnarray}\label{eq:deltaT}
\delta{T}{^0_0}&=&\frac{1}{a^2}\left[(\dot{\Phi}_0\dot{\delta\Phi}^{\ast}+\dot{\Phi}^{\ast}_0\dot{\delta\Phi})-2\psi\dot{\Phi}_0\dot{\Phi}^{\ast}_0+a^2\delta V \right]\nonumber\\
&+&\frac{iq}{a^2}\left(\Phi_0\dot\Phi^{\ast}_0-\Phi^{\ast}_0\dot\Phi_0\right)B_0,\label{eq:deltaT00}\\
\delta{T}{^0_j}&=&\frac{1}{a^2}\left(\dot\Phi_0\delta{\Phi}^{\ast}_{,j}+\dot{\Phi}^{\ast}_0 \delta\Phi_{,j}\right)
\nonumber\\
&+&\frac{iq}{a^2}\left(\Phi_0\dot\Phi^{\ast}_0-\Phi^{\ast}_0\dot\Phi_0\right)B_j,\label{eq:deltaT0j}\\
\delta{T}{^i_j}&=&-\frac{1}{a^2}\left[(\dot{\Phi}_0\dot{\delta\Phi}^{\ast}+\dot{\Phi}^{\ast}_0\dot{\delta\Phi})-2\psi\dot{\Phi}_0\dot{\Phi}^{\ast}_0-a^2\delta V \right]\delta^i_j\nonumber\\
&+&\frac{iq}{a^2}\left(\Phi_0\dot\Phi^{\ast}_0-\Phi^{\ast}_0\dot\Phi_0\right)B_0\delta^i_j.\label{eq:deltaTij}
\end{eqnarray}
From here we can define
\begin{eqnarray}
    \delta\rho_{\Phi}&=&-\frac{1}{a^2}\left[(\dot{\Phi}_0\dot{\delta\Phi}^{\ast}+\dot{\Phi}^{\ast}_0\dot{\delta\Phi})-2\psi\dot{\Phi}_0\dot{\Phi}^{\ast}_0+a^2\delta V \right]\nonumber\\
&+&\frac{iq}{a^2}\left(\Phi_0\dot\Phi^{\ast}_0-\Phi^{\ast}_0\dot\Phi_0\right)B_0.\label{eq:deltarho}
\end{eqnarray}
As usual, the perturbation of the potential is given by the Taylor series up to first order for the variables $\Phi$ and $\Phi^{\ast}$ around ($\Phi_0$,$\Phi^{\ast } _0$), we get
\begin{eqnarray}\label{eq:deltaV0}
V(\Phi,\Phi^{\ast})&=& V(\Phi_0,\Phi^{\ast}_0)\nonumber\\
+\left.\frac{\partial V}{\partial  \Phi}\right|_0\left(\Phi-\Phi_0\right)&+&\left.\frac{\partial V}{\partial\Phi^{\ast}}\right|_0\left(\Phi^{\ast}-\Phi^{\ast}_0\right)+\cdots\nonumber\\
=V\left(\Phi_0,\Phi^{\ast}_0\right)+\delta V+\cdots,
\end{eqnarray}
where $|_0=|_{(\Phi_0,\Phi_0^{\ast})}$, $\delta\Phi=\Phi-\Phi_0$ and $\delta V=V-V|_0$. Explicitly we have
\begin{equation}\label{eq:deltaV}
    \delta V=V,_{\Phi_0} \delta\Phi + V,_{\Phi_0^{\ast}}\delta\Phi^{\ast}.
\end{equation}

In the same way, we can derive the corresponding derivative
\begin{equation}
    \frac{dV}{d\Phi^\ast}=
    V_{,\Phi^\ast}|_0+ V_{,\Phi\Phi^\ast}|_0\delta\Phi
    + V_{,\Phi^\ast\Phi^\ast}|_0\delta\Phi^\ast
\end{equation}
for the scalar field potential.

Observe that if we compare the equations (\ref{eq:deltaGij}) and (\ref{eq:deltaTij}) for $i\not= j$, we see that $\left(\phi-\psi\right)^ {,i}_{,j} =0$. The simplest solution to this constraint is $\psi=\phi$.

Therefore, the Einstein's equation for the perturbed part of the scalar field reads
\begin{eqnarray}
2\nabla^2\phi-6H(\dot{\phi}+H\phi)&=&\nonumber\\
\kappa^2\left[(\dot{\Phi}_0\dot{\delta\Phi}^{\ast}+\dot{\Phi}^{\ast}_0\dot{\delta\Phi})-2\phi\dot{\Phi}_0\dot{\Phi}^{\ast}_0+a^2\delta V \right]&&\nonumber\\
+iq\kappa^2\left(\Phi_0\dot\Phi^{\ast}_0-\Phi^{\ast}_0\dot\Phi_0\right)B_0,&&\label{eq:ecuacionesdeeinstein1}\\
2(\dot{\phi}+H\phi)_{,j}=\kappa^2\left(\dot\Phi_0\delta{\Phi}^{\ast}_{,j}+\dot{\Phi}^{\ast}_0 \delta\Phi_{,j}\right)&&
\nonumber\\
+iq\kappa^2\left(\Phi_0\dot\Phi^{\ast}_0-\Phi^{\ast}_0\dot\Phi_0\right)B_j,&&\label{eq:ecuacionesdeeinstein2}\\
2\left(\ddot{\phi}+3H\dot{\phi}+(2\dot{H}+H^2)\phi\right)&=&\nonumber\\
\kappa^2\left[(\dot{\Phi}_0\dot{\delta\Phi}^{\ast}+\dot{\Phi}^{\ast}_0\dot{\delta\Phi})-2\phi\dot{\Phi}_0\dot{\Phi}^{\ast}_0-a^2\delta V \right]&&\nonumber\\
+iq\kappa^2\left(\Phi_0\dot\Phi^{\ast}_0-\Phi^{\ast}_0\dot\Phi_0\right)B_0.&&\label{eq:ecuacionesdeeinstein3}
\end{eqnarray}

Then, the field equations for the perturbations $\phi$, $\psi$ and $B_j$ are the equations (\ref{eq:ecuacionesdeeinstein1}), (\ref{eq:ecuacionesdeeinstein2}) using the equation (\ref{eq:deltaV}).

As in the case of the real scalar field (\cite{Magana:2012xe}), if we subtract the equation (\ref{eq:ecuacionesdeeinstein3}) from (\ref{eq:ecuacionesdeeinstein1}), we obtain a unique equation for the gravitational potential $\phi$
\begin{equation}\label{eq:phi}
\ddot{\phi}+6H\dot{\phi}+2\phi\left(\dot{H}+2H^2\right)-\frac{1}{a^2}\nabla^2\phi+\kappa^2\delta V\,=\,0.
\end{equation}

For the perturbation of the scalar field $\delta\Phi$ the KG equation takes the form
\begin{eqnarray}\label{eq:KGperturbada}
&&\nabla^2\delta\Phi-\ddot{\delta\Phi}-2H\dot{\delta\Phi}-\frac{iq}{a^2}(a^2\Phi_0B_0\dot)\\
&&+4\dot{\phi}\dot\Phi_0+V_{,\Phi_0\Phi_0^{*}}a^2\delta\Phi+V,_{\Phi^{*}_0\Phi^{*}_0}a^2\delta\Phi^{*}-2V,_{\Phi^{*}_0}a^2\phi=0.\nonumber
\end{eqnarray}
where we have used the KG equation in the background (\ref{eq:KGhomogenea}). This last equation completes the system of field equations to be solved.

\section{The Madelung transformation for KG fluctuations}\label{sec:4medulung}

In this section we apply the Madelung transformation to the perturbed equation (\ref{eq:ecuacionesdeeinstein1})-(\ref{eq:ecuacionesdeeinstein2}) and (\ref{eq:KGperturbada}). The goal of this is to separate the perturbation into two parts, one function that can be interpreted as number density and the other as the velocity of the scalar field as fluid. We introduce an ansatz that allows us to give exact solutions in some cases. These solutions are not necessarily physical, but we can deduce some physical aspects of them, as we will see next.

As usual, to find the shape of the perturbations $\delta\Phi$ we expand $\Phi_0$, using the polar representation of a
complex function $\Phi=R\,e^{i\theta}$, where $R=||\Phi||$ is the norm and $\theta$ the phase of the scalar field, we have
\begin{equation}
\Phi=\Phi_0+\left.\frac{\partial\Phi}{\partial R}\right|_0(R-R_0)+\left.\frac{\partial \Phi}{\partial \theta}\right|_0(\theta-\theta_0)+\cdots,
\end{equation}
where $|_0=|_{(R_0,\theta_0)}$. Thus, we obtain
\begin{equation}\label{eq:deltaPhi1}
\delta\Phi= \left(\frac{\delta R}{R_0}+i\delta\theta\right)\,\Phi_0+\cdots\,\,.
\end{equation}

We will define the following quantities as ansatz to simplify the field equations, which will help us reduce the equations to quadratures. Let be {
\begin{eqnarray}\label{eq:deltaPhi2}
\Phi_0&=&\sqrt{n_0}\,e^{i\theta_0}=\sqrt{n_0}\,e^{i(S_0-\omega_0 \eta)},\\
\delta\Phi&=&\sqrt{n}\,e^{i(S_0+\delta S-\omega_0 \eta)},\label{eq:deltaPhi3}
\end{eqnarray}}
 {The physical interpretation for $n_0$, $n$, $S_0$, $\delta S$ and $S$ could be controversial, but here this transformation is used only as ansatz to analytically solve the equations.}
 If we compare (\ref{eq:deltaPhi1}), (\ref{eq:deltaPhi2}) with (\ref{eq:deltaPhi3}) we can see that $R^2_0=n_0$, $\delta R^2=n$ and $\delta\theta=\sqrt{n/n_0}\delta S$. We will take perturbations where if $\phi$ is of the order $\epsilon$, then $\delta\Phi\, , \delta R$ and $\delta S$ are also of this order, while $\delta\theta$ and $n $ are perturbations of order $\epsilon^2$. We will neglect perturbations of order $\epsilon^2$ and beyond.
 
    With this in mind we can rewrite the equation (\ref{eq:G00L}) in terms of these definitions, we get
  \begin{equation}\label{eq:rhoPhin0}
      \rho_{\Phi_0}=\frac{1}{4}\frac{\dot n_0^2}{n_0}+n_0\dot\theta_0^2+(m^2+\lambda n_0)n_0\,a^2,
  \end{equation}
  while, equations (\ref{eq:ecuacionesdeeinstein1})-(\ref{eq:ecuacionesdeeinstein3}) written in terms of the definitions (\ref{eq:deltaPhi2}) and (\ref{eq:deltaPhi3}), become
\begin{eqnarray}
&&2\nabla^2\phi-6H(\dot{\phi}+H\phi)=\nonumber\\
&&\kappa^2\left[\frac{\dot n_0}{2\sqrt{n_0}}\frac{\dot n}{\sqrt{n}}+2\sqrt{n_0\,n}\dot\theta_0^2-\left(2n_0\dot\theta_0^2+\frac{1}{2}\frac{\dot n_0^2}{ n_0}\right)\phi \right]\nonumber\\
&&+\kappa^2[2qn_0\dot\theta_0^2B_0+a^2\delta V],\label{eq:ecuacionesdeeinstein1-n}\\
&&2(\dot{\phi}+H\phi)_{,j}=\kappa^2\left(\frac{\dot n_0}{2\sqrt{n_0}}\frac{n_{,j}}{\sqrt{n}}
+2qn_0\,\dot\theta_0B_j\right),\label{eq:ecuacionesdeeinstein2-n}\\
&&2\left(\ddot{\phi}+3H\dot{\phi}+(2\dot{H}+H^2)\phi\right)=\nonumber\\
&&\kappa^2\left[\frac{\dot n_0}{2\sqrt{n_0}}\frac{\dot n}{\sqrt{n}}+2\sqrt{n_0\,n}\dot\theta_0^2-\left(2n_0\dot\theta_0^2+\frac{1}{2}\frac{\dot n_0^2}{ n_0}\right)\phi\right]\nonumber\\
&&+\kappa^2[2q n_0\dot\theta_0^2B_0-a^2\delta V ].\label{eq:ecuacionesdeeinstein3-n}
\end{eqnarray}

In terms of the new variables, the perturbed scalar field potential $\delta V$ reads
\begin{equation}
    \delta V=2(m^2+\lambda n_0)\sqrt{n_0\,n}.
\end{equation}
With the same definitions, the KG equation (\ref{eq:KG}) is separated into its real and imaginary parts. For the background we get
\begin{eqnarray}
\frac{1}{a^2}\left(\frac{a^2\dot n_0}{\sqrt{n_0}}\right)^{\dot{}}-2(\sqrt{n_0}\dot\theta_0^2-V_{,\Phi_0}a^2e^{-i\theta_0})&=&0,\label{eq:KG0Re-n}\\
\left(a^2n_0\dot\theta_0\right)^{\dot{}}&=&0,\label{eq:KG0Im-n}
\end{eqnarray}
while, for the perturbation, we obtain
\begin{eqnarray}
&&\left[\frac{1}{a^2}\left(\frac{a^2\dot n_0}{\sqrt{n_0}}\right)^{\dot{}}-2\sqrt{n_0}\dot\theta_0^2\right]\phi-\frac{1}{2a^2}\left(\frac{a^2\dot n}{\sqrt{n}}\right)^{\dot{}}\\
&+&(\dot\theta_0^2+m^2+3\lambda n_0)\sqrt{n}+\frac{2\,\dot n_0}{\sqrt{n_0}}\dot\phi+
\sqrt{n_0}\dot\theta_0B_0+\frac{\Box\sqrt{n}}{\sqrt{n}}=0, \label{eq:KGRe-n}\nonumber\\
&&\frac{1}{\sqrt{n}}(a^2\,n\,\dot\theta_0)^{\dot{}}-\frac{1}{\phi\,\sqrt{n_0}}(a^2\,n_0\phi^2\dot\theta_0\dot)+\frac{1}{2}(a^2\sqrt{n_0}\,B_0\dot)=0,\nonumber\\\label{eq:KGIm-n}
\end{eqnarray}
for the real and imaginary parts, respectively. 

We will use the standard definition for density contrast $\delta$, which is
 \begin{equation}
     \delta=\frac{\delta\Phi}{\Phi_0}\approx\sqrt{\frac{n}{n_0}}.\label{eq:delta}
 \end{equation}
It is convenient to use the equation (\ref{eq:KG0Im-n}) and the density contrast definition (\ref{eq:delta}) in the perturbed equations. With this in mind, the equations (\ref{eq:ecuacionesdeeinstein1-n}) and (\ref{eq:ecuacionesdeeinstein2-n}) transform into
\begin{eqnarray}
&&2\nabla^2\phi-6H(\dot{\phi}+H\phi)=\label{eq:ecuacionesdeeinstein1-d}\\
&&\kappa^2\left[\dot n_0\dot\delta+\left(\frac{1}{2}\frac{\dot n_0^2}{n_0}+\frac{2s_0^2}{a^4\,n_0}\right)(\delta-\phi)+a^2\delta V +2q\frac{s_0}{a^2}B_0\right],\nonumber
\\
&&2(\dot{\phi}+H\phi)_{,j}=\kappa^2\left(\dot n_0\delta_{,j}+\frac{2q\,s_0}{a^2}B_j\right).\label{eq:ecuacionesdeeinstein2-d}
\end{eqnarray}
On the other hand, the imaginary part of the KG equation (\ref{eq:KGIm-n}) becomes
\begin{equation}\label{eq:KGIm-d}
    \dot\delta-2\dot\phi+\frac{\sqrt{n_0}}{2}\left(\frac{a^2\sqrt{n_0}}{s_0}B_0\right)^{\dot{}}=0,
\end{equation}
where we have used the equation (\ref{eq:KG0Im-n}) and stated that $a^2n_0\dot\theta_0=s_0$, where $s_0$ is a constant.

Maxwell's equation (\ref{eq:Maxwell}) reduces to
\begin{eqnarray}\label{eq:Maxwell-d}
   \eta^{\nu\alpha}B_{\mu\nu,\alpha}&=&\frac{2a^2n_0}{m^2}(qB_\mu+\dot\theta_0\delta\delta^0_\mu)\nonumber\\
   &=&\frac{2a^2n_0}{m^2}qB_\mu+\frac{2s_0}{m^2}\delta\delta^0_\mu
\end{eqnarray}
The electromagnetic energy stress tensor is quadratic in $B_\mu$ and therefore has no contribution to the Einstein equations.

\section{The Fourier space}\label{sec:5Fourier}

In this section we rewrite the field equations in Fourier space, where the perturbations will become
\begin{equation}
       f=\int f_k \,e^{ik_i x^i}d^3k
\end{equation}
for any perturbation $f$ where, as usual, $k$ is the wavenumber, being $k=2\pi/\lambda_k$ where $\lambda_k$ is the length scale of the perturbation. In Fourier space, the equations (\ref{eq:ecuacionesdeeinstein1})-(\ref{eq:ecuacionesdeeinstein3}) reduce to
\begin{eqnarray}
-2k^2\phi_k-6H(\dot{\phi}_k+H\phi_k)&=&\nonumber\\
\kappa^2\left[(\dot{\Phi}_0\dot{\delta\Phi}_k^{\ast}+\dot{\Phi}^{\ast}_0\dot{\delta\Phi}_k)-2\psi_k\dot{\Phi}_0\dot{\Phi}^{\ast}_0+a^2\delta V_k \right]&&\nonumber\\
+iq\kappa^2\left(\Phi_0\dot\Phi^{\ast}_0-\Phi^{\ast}_0\dot\Phi_0\right)B_{0k},&&\label{eq:ecuacionesdeeinstein1-F}\\
2(\dot{\phi}_k+H\phi_k)=\kappa^2\left(\dot\Phi_0\delta{\Phi}_k^{\ast}+\dot{\Phi}^{\ast}_0 \delta\Phi_{k}\right)&&
\nonumber\\
+q\kappa^2\left(\Phi_0\dot\Phi^{\ast}_0-\Phi^{\ast}_0\dot\Phi_0\right)\frac{B_{jk}}{k_j},&&\label{eq:ecuacionesdeeinstein2-F}\\
2\left(\ddot{\phi}_k+3H\dot{\phi}_k+(2\dot{H}+H^2)\phi_k\right)&=&\nonumber\\
\kappa^2\left[(\dot{\Phi}_0\dot{\delta\Phi}_k^{\ast}+\dot{\Phi}^{\ast}_0\dot{\delta\Phi}_k)-2\psi_k\dot{\Phi}_0\dot{\Phi}^{\ast}_0-a^2\delta V_k \right]&&\nonumber\\
+iq\kappa^2\left(\Phi_0\dot\Phi^{\ast}_0-\Phi^{\ast}_0\dot\Phi_0\right)B_{0k}.&&\label{eq:ecuacionesdeeinstein3-F}
\end{eqnarray}
In the same way the KG equation  (\ref{eq:KGperturbada}) in Fourier space becomes
\begin{eqnarray}\label{eq:KGperturbada-F}
&&-k^2\delta\Phi_k-\ddot{\delta\Phi}_k-2H\dot{\delta\Phi}_k-\frac{iq}{a^2}(a^2\Phi_0B_{0k}\dot)\\
&&+4\dot{\phi}_k\dot{\Phi_0}+V_{,\Phi_0\Phi_0^{*}}a^2\delta\Phi_k+V_{,\Phi^{*}_0\Phi^{*}_0}a^2\delta\Phi_k^{*}-2V_{,\Phi^{*}_0}a^2\phi_k=0.\nonumber
\end{eqnarray}

It is more convenient to write this last equation in terms of the density contrast $\delta$, we can set in (\ref{eq:KGperturbada-F}) that $\delta\Phi=\Phi_0\delta$ to obtain
\begin{eqnarray}\label{eq:KG-dk}
    &&\ddot\delta_k+2\left(H+\frac{\dot\Phi_0}{\Phi_0}\right)\dot\delta_k-
    \left(V_{,\Phi_0\Phi_0^\ast}+\frac{V_{,\Phi_0^\ast\Phi_0^\ast}-V,_{\Phi_0}}{\Phi_0}\right)a^2\delta_k\nonumber\\
   && +k^2\delta_k+k_{j}B_k^j+i(\dot B_{0k}+2HB_{0k})\nonumber\\
    &&-4\left({\dot\phi}_k+H\phi_k-\frac{1}{4}B_{0k}\right)\frac{\dot\Phi_0}{\Phi_0}\nonumber\\
    &&+2\left(2H\frac{\dot\Phi_0}{\Phi_0}+\frac{V_{,\Phi_0}}{\Phi_0}a^2\right)\phi_k=0
\end{eqnarray}
Similarly, the equation for the Newtonian potential $\phi$ in Fourier space reads
\begin{equation}\label{eq:phik}
\ddot{\phi_k}+6H\dot{\phi_k}+2\phi_k\left(\dot{H}+2H^2\right)+\frac{k^2}{a^2}\phi_k+\kappa^2\delta V_k\,=\,0,
\end{equation}
where $\delta V_k=(V,_{\Phi_0} \Phi_0 + V,_{\Phi_0^{\ast}}\Phi_0^{\ast})\delta_k $

The equations (\ref{eq:ecuacionesdeeinstein1-F})-(\ref{eq:ecuacionesdeeinstein3-F}), the KG equation (\ref{eq:KG-dk}) along with Maxwell's equations (\ref {eq:Maxwell-d}) form a complete system of linear differential equations in the variable $\eta$ that can be solved numerically. We can proceed in the following two ways. We can solve the system by finding reasonable initial conditions for the system, where we know nothing about the electromagnetic field. We can guess the behavior and say something about it. The second way is the following. We know that for the development of fluctuations the electromagnetic field plays a non-essential role, the fluctuations are dominated by the gravitational field. In this case, then, we can solve the system without taking Maxwell's equations into account, neglecting the electromagnetic field due to the evolution of the fluctuations and then, with the same equations, evaluate the behavior of the electromagnetic field. Before doing this, we will take two approximate situations of the system. 

\section{A toy model without electromagnetic field}\label{sec:6Bigual0}
 \begin{figure}
\centering
\includegraphics[width=0.5\textwidth]{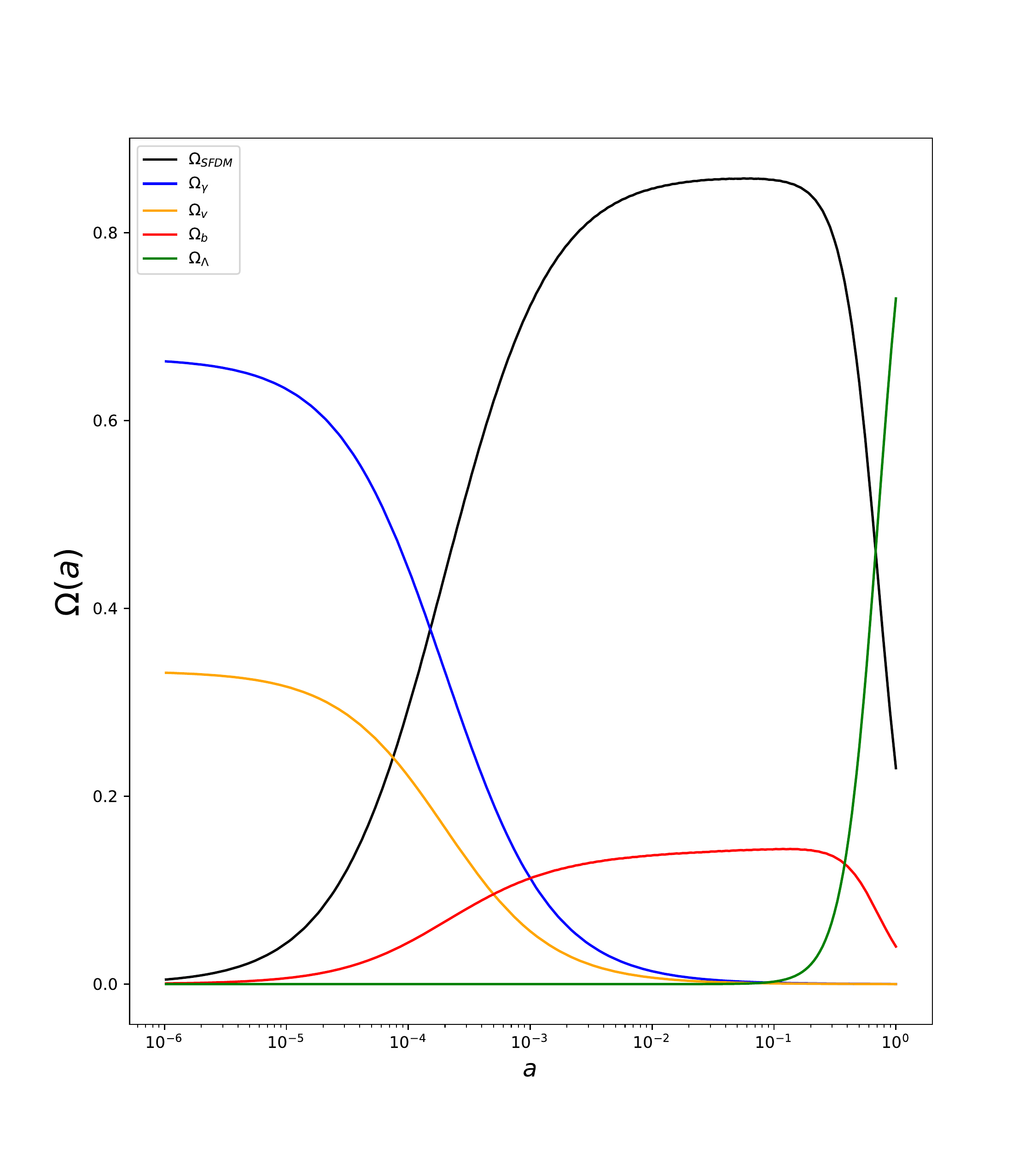}
\caption{The evolution of the density rates $\Omega_{SFDM}$ (black line), $\Omega_b$ (red line), $\Omega_r$ (blue line), $\Omega_\nu$ (yellow line) and $\Omega_\Lambda$ (green line) for the SFDM model. The evolution is almost exactly the same in the LCDM model. This behavior was first introduced in Matos\&Urena-Lopez (2001) }
\label{fig:sfdm_Complejo}
\end{figure}
 \begin{figure}
\centering
\includegraphics[width=0.4\textwidth]{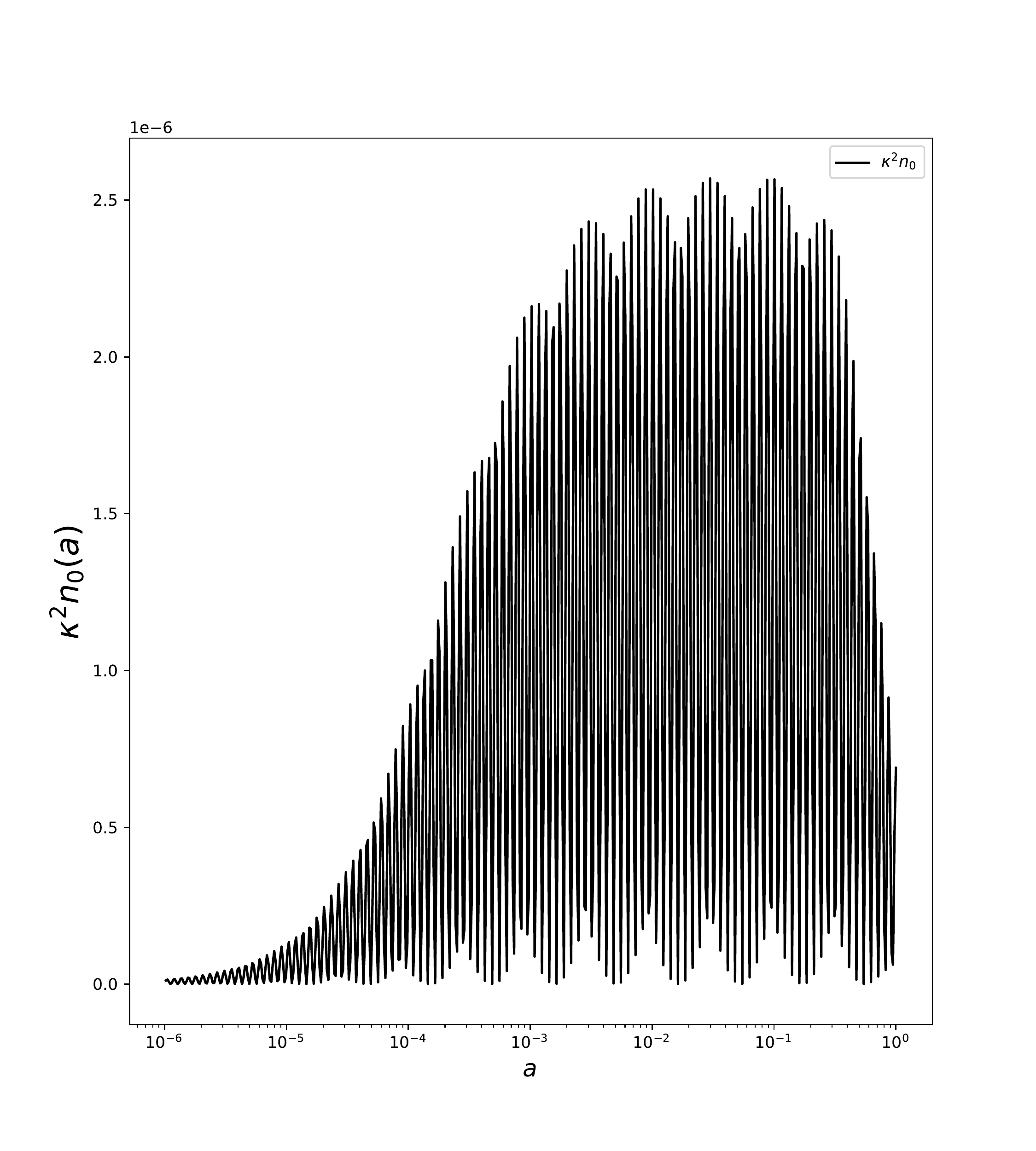}
\includegraphics[width=0.4\textwidth]{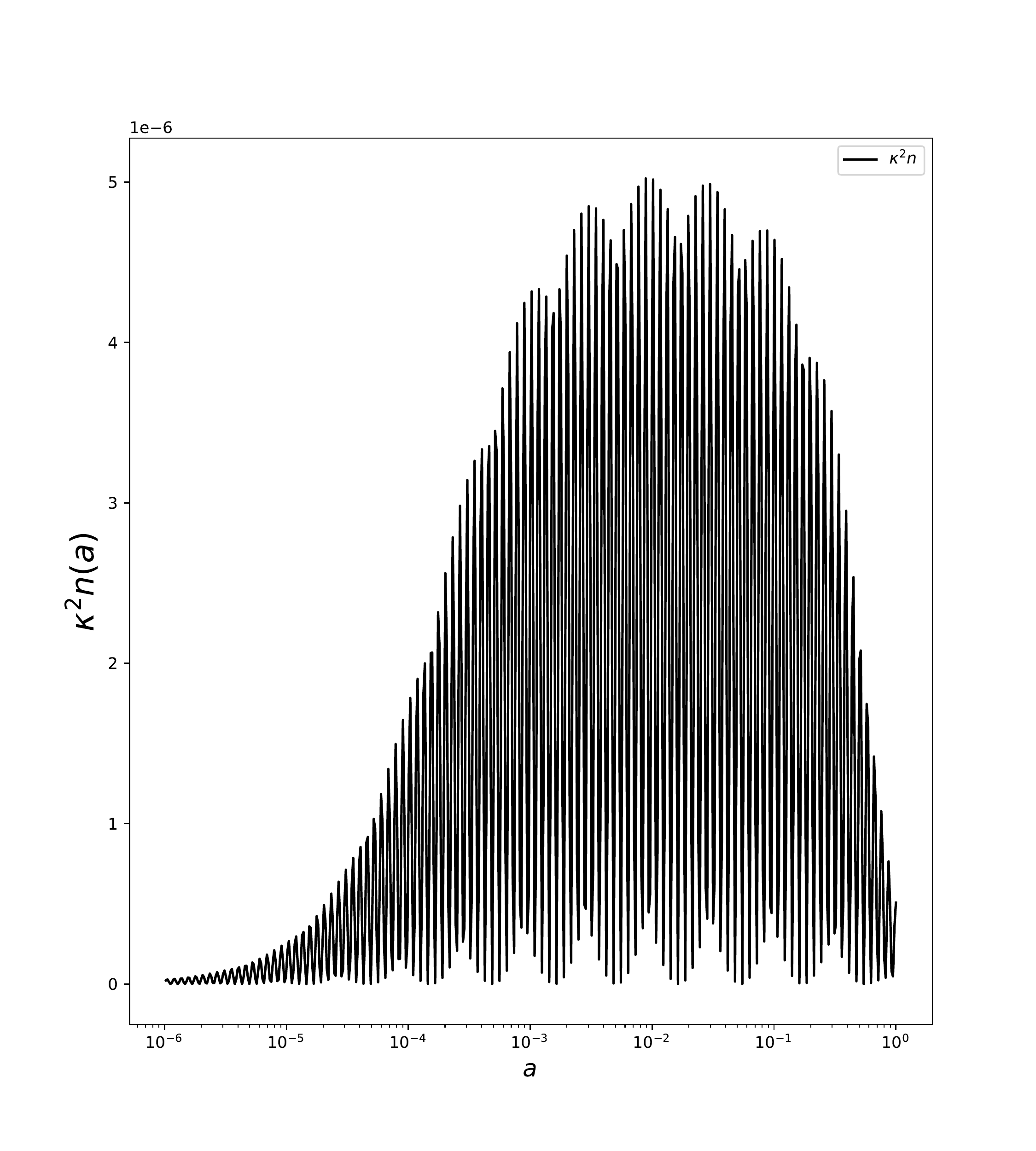}
\caption{Upper panel: the evolution of the background density rate $n_0$. The oscillations are the unmistakable imprint of the quantum behavior of density, when combined with the evolution of the potential energy $\phi_0$ the result is the line observed in Fig.\ref{fig:sfdm_Complejo} for the SFDM. Lower panel: the evolution of the fluctuation density rate $n$. Here the evolution is very similar to $n_0$, due to the approximation we are using in this example.}
\label{fig:Conqceron0n}
\end{figure}
 \begin{figure}
\centering
\includegraphics[width=0.4\textwidth]{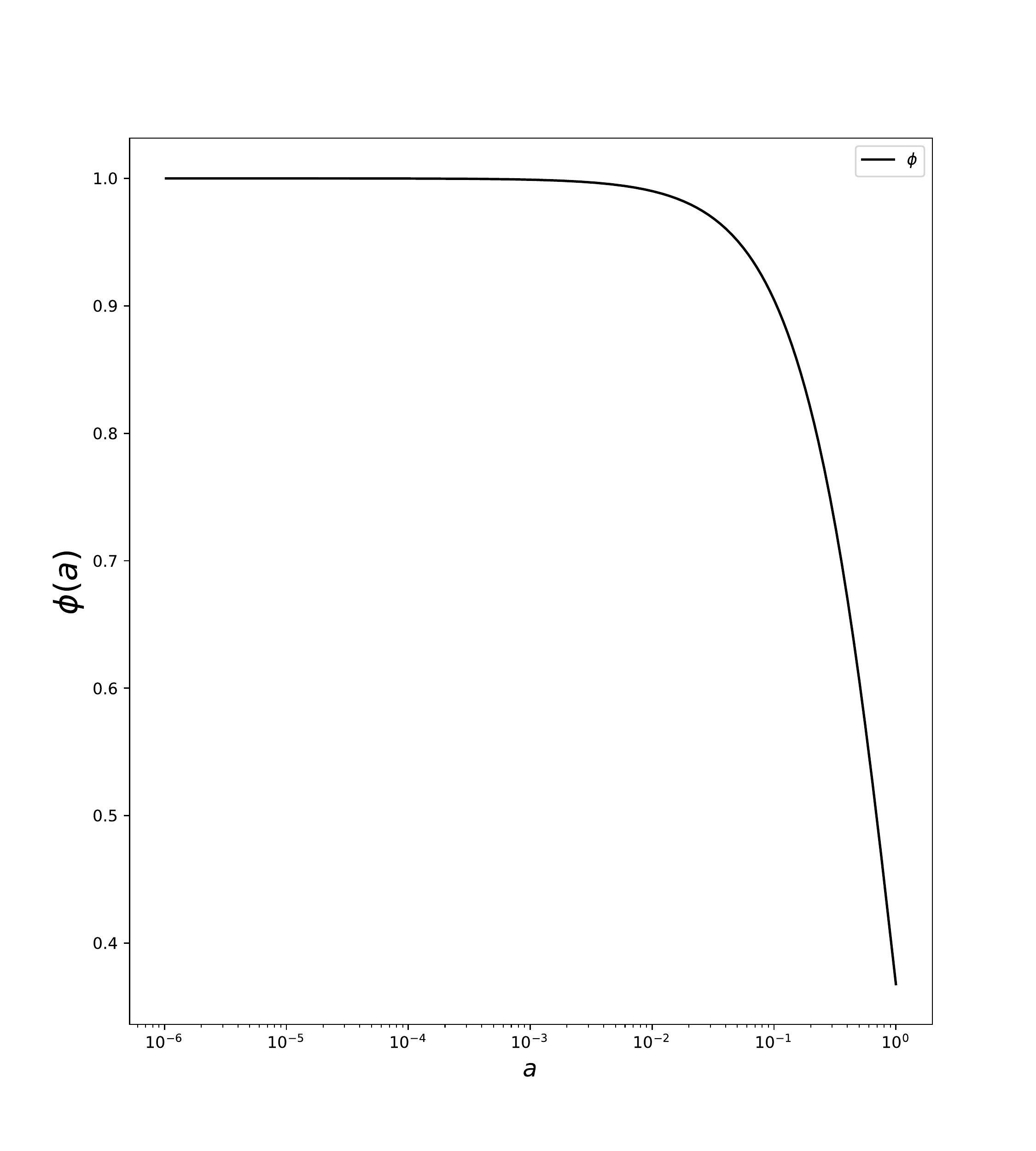}
\includegraphics[width=0.4\textwidth]{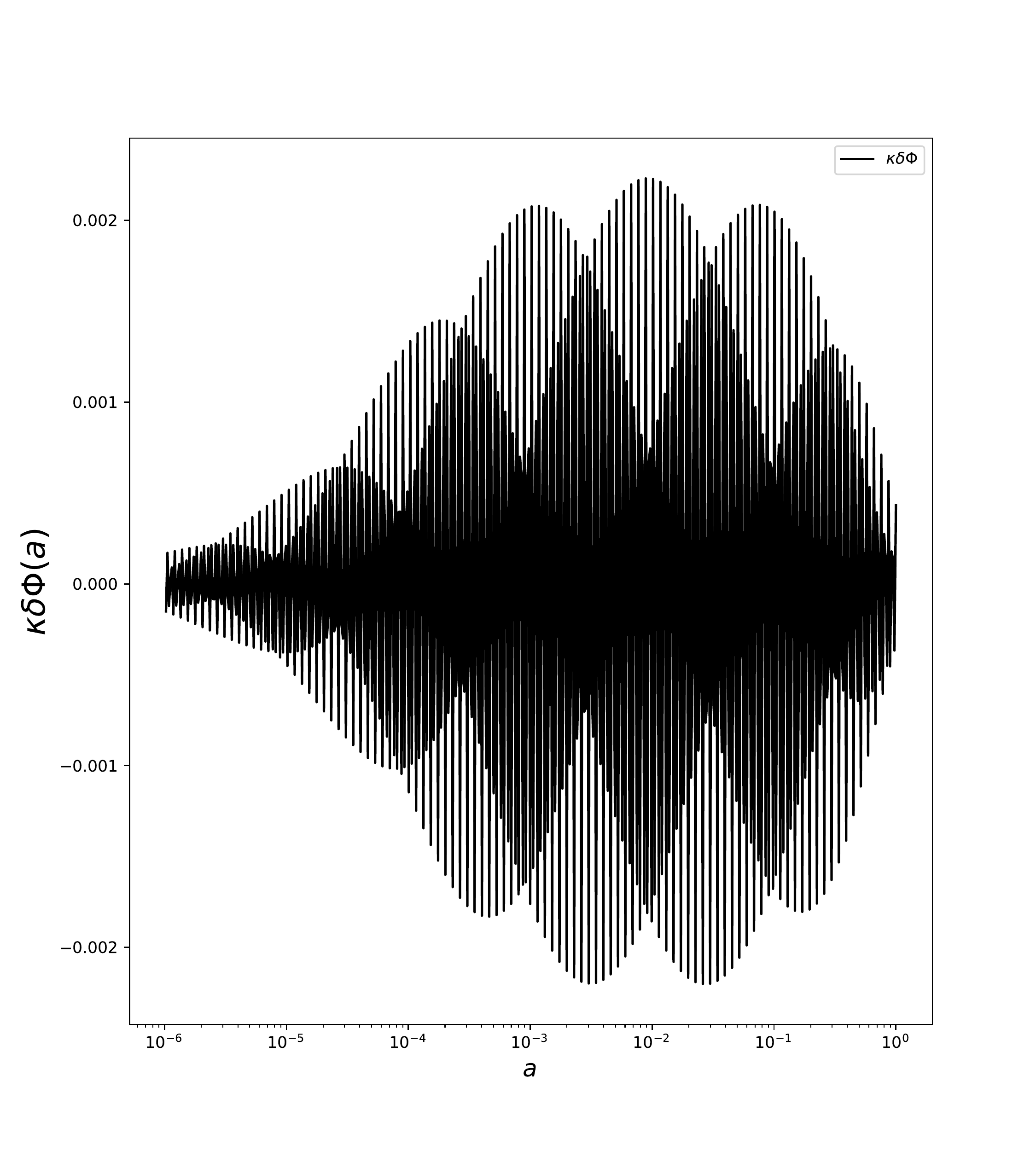}
\caption{In this figure we show the evolution of the Newtonian potential $\phi$ (upper panel) and the fluctuation of the scalar field $\delta\Phi$ (lower panel) for the exact solution (\ref{eq:solphi}). Here we solve the dynamical system for $n_0$ and $a$ and with the solution (\ref{eq:solphi}) we plot the Newtonian potential $\phi$ and the fluctuation of the scalar field $\delta\Phi$ of the equation ( \ref{eq:soldeltaPhi}). Note that the Newtonian potential evolves smoothly during the history of the universe, while the scalar field fluctuation has strong oscillations as a consequence of its quantum nature. This solution is just a toy model. }
\label{fig:Conqcerophidelphi}
\end{figure}
\begin{figure}
\centering
\includegraphics[width=0.50\textwidth]{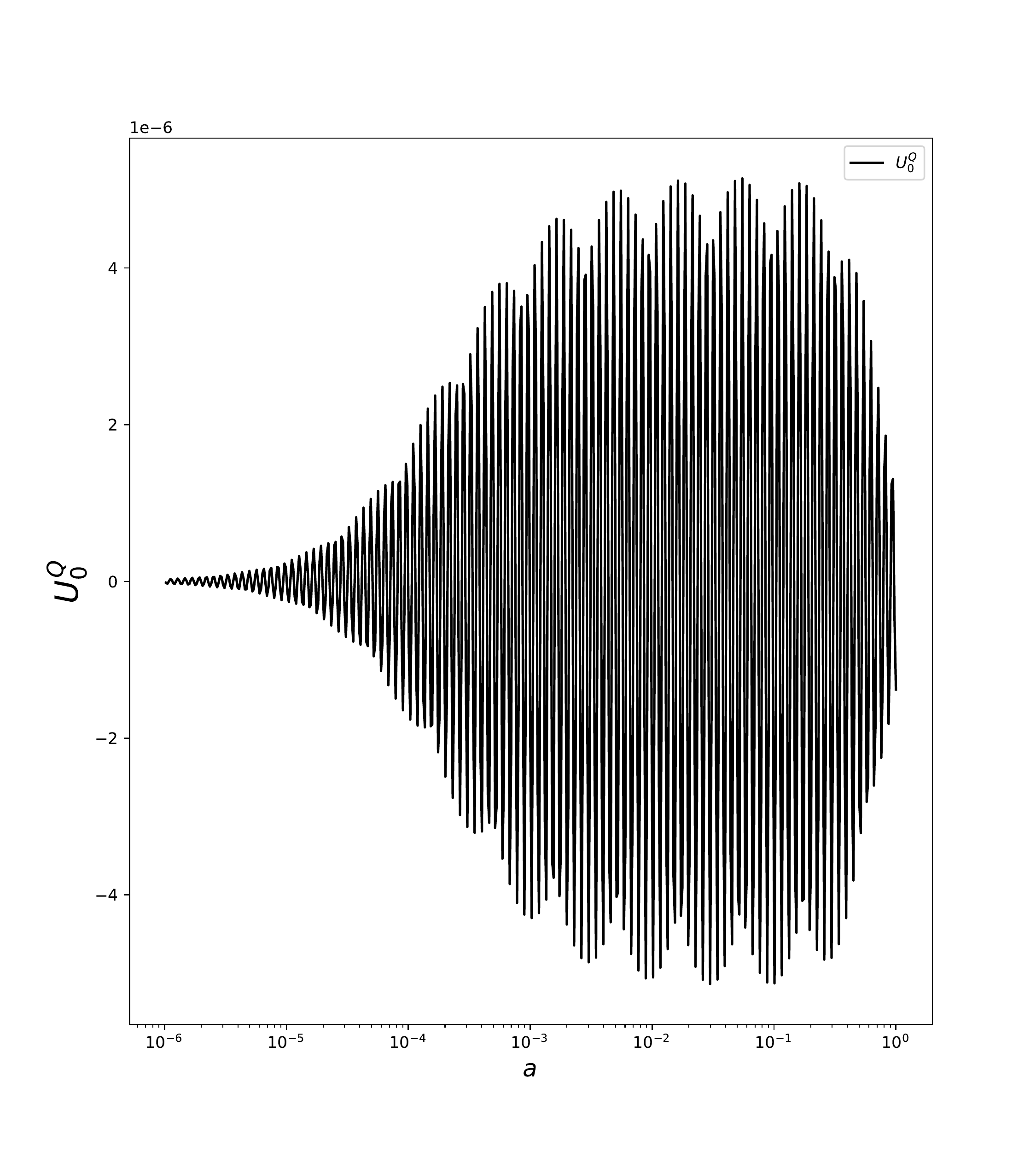}
\caption{The quantum potential $U_0^Q$ defined in (\ref{eq:UQ}) for the SFDM model. Here the quantum potential $U^Q$ is plotted for the cosmological background. This is the main difference between the classical hydrodynamic behavior of the DM, such as LCDM, and the SFDM model. The regions where $U^Q\sim 0$, the SFDM behaves like a fluid, but the regions where it is not, the quantum characteristic of the SFDM is important and differs from the hydrodynamics.  }
\label{fig:ConqceroU0Q}
\end{figure}

In this section we integrate some of the equations that we have derived in terms of the variables $a$ and $n_0$ for the case that the electromagnetic field disappears. Note that the equation (\ref{eq:KGIm-d}) implies that $\delta=2\phi+X(x^j)$, {where $X$ is an arbitrary function}. We start with a toy model that integrates the system assuming $X=0$. As we will see, this choice of the function $X$ does not lead to realistic results, but it allows us to integrate the system very easily and gives us an insight into the realistic results. If we do so, the equation (\ref{eq:ecuacionesdeeinstein2-d}) can be integrated to give
\begin{equation}\label{eq:solphi}
    a\phi=P_0e^{\kappa^2\,n_0},
\end{equation}
where $P_0=P_0(x^i)$ {is an arbitrary function}. Now, using the equation (\ref{eq:ecuacionesdeeinstein1-d}) we can substitute the two previous results and obtain
\begin{equation}\label{eq:P0}
    \nabla^2P_0+\Omega^2P_0=0,
\end{equation}
where 
\begin{eqnarray}
    \Omega^2&=&-\kappa^2\left[\kappa^2\dot n_0^2+\frac{2\dot a\dot n_0}{a}+\left(\frac{1}{4}\frac{\dot n_0^2}{n_0}+\frac{s_0^2}{a^4\,n_0}\right)\right.\nonumber\\
    &+&\left.2a^2(m^2+\lambda n_0)n_0\right].\nonumber
\end{eqnarray}

The simplest solutions of this equation are $P_0=P_{lj}RY_l^j$, where $P_{lj}$ are constants, $R=R(r)$ and $Y_l^j(\theta,\varphi) $ are the spherical harmonic polynomials. With this ansatz this equation reduces to
\begin{equation}
    \frac{d}{dr}\left(r^2\frac{dR}{dr}\right)+(\Omega^2r^2-l(l+1))R=0.
\end{equation}
The simplest solution of the equation (\ref{eq:P0}) is for $l=0$, this is
\begin{equation}\label{eq:solP0}
    P_0=P_{00}\frac{\sin(\Omega r)}{r},
\end{equation}
where $P_{00}$ is a constant and $r$ is the radial spherical coordinate, $r^2=x^2+y^2+z^2$ is the radius of a sphere. The following solutions in degree of difficulty are the solutions with $l=1$. {The density of the fluctuation that collapses to form the structure is $\delta\Phi\delta\Phi^*$. In this case, using the equations (\ref{eq:delta}), (\ref{eq:solphi}) and the previous results we obtain
\begin{equation}
    \delta\Phi\delta\Phi^*=4|P_{lj}|^2R^2Y^j_lY^{*j}_l\frac{\Phi_0\Phi_0^*e^{2\kappa^2n_0}}{a^2}.
\end{equation}}
{However, the spherical harmonic functions $Y^j_lY^{*j}_l$ for $l=1$ contain two bubbles, one north and one south of the origin of coordinates. These bubbles are a direct consequence of the quantum nature of the scalar field.}
These solutions are very interesting, since they are the $p$ states of an atom. In this simple case, these solutions are obtained directly as an exact solution. However, in \cite{Guzman:2022vxl} it will be shown that in the physical case, solved with numerical methods, these solutions are very similar to the numerical ones and this $p$ state behavior of an atom is practically the same. This form of dark matter can explain the anomalous trajectories observed in satellite galaxies, the VPO, and the FB observed in our galaxy.

Now we can know the complete behavior of the fluctuation $\delta\Phi$, as $\delta=2(P_0/a)\,e^{\kappa^2 n_0}$, we obtain that
\begin{equation}\label{eq:soldeltaPhi}
    \delta\Phi=2P_{00}\frac{\sin(\Omega r)}{r}\frac{e^{\kappa^2 n_0}}{a}\Phi_0.
\end{equation}

To obtain a solution of the quantities $\delta\Phi$, $\phi$ and $\delta$ we have to integrate the equations (\ref{eq:G00L}) and (\ref{eq:KGhomogenea}) together with the rest of equations for the components of baryons, radiation, dark energy and curvature.

In what follows we solve the field equations using a numerical code. To do so, we first have to define variables without units. As background we have that the complete system that we have to solve is then
\begin{eqnarray}\label{eq:HT}
    \mathcal{H}^2&=&\frac{\kappa^2}{3}\left(\frac{\dot\Phi_0\dot\Phi_0^\ast}{a^2}+V+\rho_b+\rho_r+\rho_\nu+\rho_\Lambda\right),
\end{eqnarray}
where $b$ is for baryons, $z$ for radiation, $\nu$ for neutrinos and $\rho_\Lambda$ for the cosmological constant which here is dark energy.

It is convenient to derive the equation (\ref{eq:HT}) to have two pair differential equations instead of one. We obtain
\begin{eqnarray}\label{eq:HTp}
     \mathcal{\dot H}&=&-\frac{\kappa^2}{2}\left(2\frac{\dot\Phi_0\dot\Phi_0^\ast}{a^2}+\rho_b+\frac{4}{3}\rho_r+\frac{4}{3}\rho_\nu\right),
\end{eqnarray}
where $\mathcal{H}=H/a$. The continuity equations for the components of the universe are as usual $\dot\rho_X=-3H(1+\omega_X)\rho_X$

It is convenient to change the variable $\eta$ to the e-folding variable $N=ln(a)$.
Note that the scalar field density can also be written in terms of $n_0$ and $N$ as
\begin{eqnarray}\label{eq:rho-N}
    \rho_{\Phi_0}=H^2\left(\frac{ n_0'^2}{4n_0}e^{-2N}+\frac{s_ 0^2}{n_0}\frac{e^{-6N}}{H^2} +(m^2+\lambda n_0)\frac{n_0}{H^2} \right),
\end{eqnarray}
where now a prime means a derivative with respect to $N$, that is, ${}{'} = d/dN$.
We now write the KG equation in the background, this equation in terms of the new variables reads
\begin{eqnarray}\label{eq:KG-N}
    &&n_0''+\frac{H'}{H}n_0'-\frac{1}{2}\frac{n_0'^2}{n_0}+2n'_0\nonumber\\
    &-&\frac{2s_0^2}{n_0}\frac{e^{-4N}}{H^2}+2(m^2+\lambda n_0)n_0\frac{e^{2N}}{H^2}=0.
\end{eqnarray}

Using the equation (\ref{eq:G00LP}), in terms of the new variables, the derivative of the scalar field density is
\begin{eqnarray}\label{eq:rhoP-N}
      \rho'_{\Phi_0}&=&-6H\left(\frac{n_0'^2}{4n_0}e^{-2N}+\frac{s_0^2}{n_0}\frac{e^{ -6N}}{H^2}\right).
\end{eqnarray}
   Note that the equations (\ref{eq:HT}) and (\ref{eq:HTp}) are a system of differential equations for $n_0$ and $H$, together with the continuity equations for each component and the equation of KG (\ref{eq:KGhomogenea}) that can be solved numerically. To do this, we define unitless quantities as follows
 \begin{eqnarray}\label{eq:defxyuv}
     x&=&\frac{\kappa}{\sqrt{3}}\frac{\dot\Phi_0}{a\mathcal{H}},\,\,\,\,\,\,\,\,y=\frac{\kappa}{\sqrt{3}}\frac{\dot\Phi^\ast_0}{a\mathcal{H}},\nonumber\\
     u&=&\frac{\kappa}{\sqrt{3}}\frac{m\Phi_0}{\mathcal{H}},\,\,\,\,\,\,\,\,v=\frac{\kappa}{\sqrt{3}}\frac{m\Phi^\ast_0}{\mathcal{H}},\nonumber\\
    b&=&\frac{\kappa}{\sqrt{3}}\frac{\sqrt{\rho_b}}{\mathcal{H}},\,\,\,\,\,\,\,\, l=\frac{\kappa}{\sqrt{3}}\frac{\sqrt{\rho_\Lambda}}{\mathcal{H}},\nonumber\\
    \nu&=&\frac{\kappa}{\sqrt{3}}\frac{\sqrt{\rho_\nu}}{\mathcal{H}},\,\,\,\,\,\,\,\, z=\frac{\kappa}{\sqrt{3}}\frac{\sqrt{\rho_z}}{\mathcal{H}},\nonumber\\
    s&=&\frac{m}{\mathcal{H}}.
 \end{eqnarray}
  In terms of these variables, the Friedmann equations and the continuity equations for each component are
 \begin{eqnarray}\label{eq:xpypupvp}
     x'&=&-3x-su+\frac{3}{2}\Pi x,\nonumber\\
     y'&=&-3y-sv+\frac{3}{2}\Pi y,\nonumber\\
     u'&=&sx+\frac{3}{2}\Pi u,\,\,\,\,\,\,\,\,
     v'=sy+\frac{3}{2}\Pi v,\nonumber\\
     b'&=&\frac{3}{2}(\Pi-1)b,\,\,\,\,\,\,\,\,l'=\frac{3}{2}\Pi l,\,\,\,\,\,\,\,\,s'=\frac{3}{2}\Pi s,
     \nonumber\\
     z'&=&\frac{3}{2}(\Pi-\frac{4}{3})z,\,\,\,\,\,\,\,\,
     \nu'=\frac{3}{2}(\Pi-\frac{4}{3})\nu,
 \end{eqnarray}
 where $\Pi$ is defined as
 \begin{equation}
     -\frac{\mathcal{\dot H} }{\mathcal{H}^2}=\frac{3}{2}(2xy+b^2+\frac{4}{3}z^2+\frac{4}{3}\nu^2)=\frac{3}{2}\Pi.
 \end{equation}

If we define, as usual, $\Omega_{X}=\rho_X/\rho_c$ as the density ratio corresponding to the density $\rho_X$, where $\rho_c=3H^2/\kappa^2$ is the critical density, $\Omega_\Lambda$ is the cosmological constant, $\Omega_{0b}\,,\Omega_{0r}\,$ and $\Omega_{0\nu}$ respectively are the value of the current density rate for the baryons, radiation and neutrinos and the scalar field density ratio is $\Omega_{\Phi_0}=\rho_{\Phi_0}/\rho_c$, in this case the equations system reads
 \begin{eqnarray}
     \Omega_{\Phi_0}&=&xy+uv, \,\,\,\,\,\,\,\,\Omega_b=b^2,\nonumber\\
     \Omega_{r}&=&z^2, \,\,\,\,\,\,\,\,\Omega_\nu=\nu^2,\,\,\,\,\,\,\,\,\Omega_\Lambda=l^2.\nonumber\\
 \end{eqnarray}
 The Friedmann equation transforms into a constraint equation given by
 \begin{equation}\label{eq:F}
     F=xy+uv+b^2+z^2+\nu^2+l^2=1.
 \end{equation}
 This last equation can be used to control the convergence of the numerical code. It can be seen that if we derive the equation (\ref{eq:F}), using equations (\ref{eq:xpypupvp}) we get
 \begin{equation}
     F'=(6xy+3b^2+4z^2+4\nu^2)(F-1),
 \end{equation}
 which is $0$ if $F=1$, that is, if the Freedman restriction (\ref{eq:F}) is satisfied.
 
 We solve equations (\ref{eq:xpypupvp}) numerically, using an Adams–Badsforth–Moulton algorithm. The results are represented in the figures. In fig.\ref{fig:sfdm_Complejo} we see the complete evolution of the density rates. We see that this evolution is very similar to the LCDM model. Fig.\ref{fig:Conqceron0n} shows the evolution of $n_0$, the scalar field density number for the background. This evolution is independent of the approximation that we are dealing with in this case, this behavior for $n_0$ will be maintained in the following cases that we will present in this work. Fig.\ref{fig:Conqceron0n} also shows the evolution of the density number $n$ for the fluctuation. Of course, this behavior of $n$ is not expected in a realistic universe, remember that for these solutions we have set $X=0$, and this is not the case in a realistic universe. This can also be seen in the evolution of the Newtonian potential $\phi$ and the perturbation $\delta\Phi$ in fig.\ref{fig:Conqcerophidelphi}. However, observe the evolution of the quantum potential $U_0^Q$ in fig.\ref{fig:ConqceroU0Q}, these oscillations are a typical characteristic of a quantum field. Again, this behavior does not depend on the approximations we are dealing with here. Note that these oscillations throughout the history of the universe are the most important feature of the SFDM model. We see that this potential, which is just the difference between a quantum field and a classical one, plays a very important role in the evolution of the universe, if the DM is of a scalar field nature.
 
 \section{A toy model with electromagnetic field}\label{sec:7Bno0}

\begin{figure}
\centering
\includegraphics[width=0.40\textwidth]{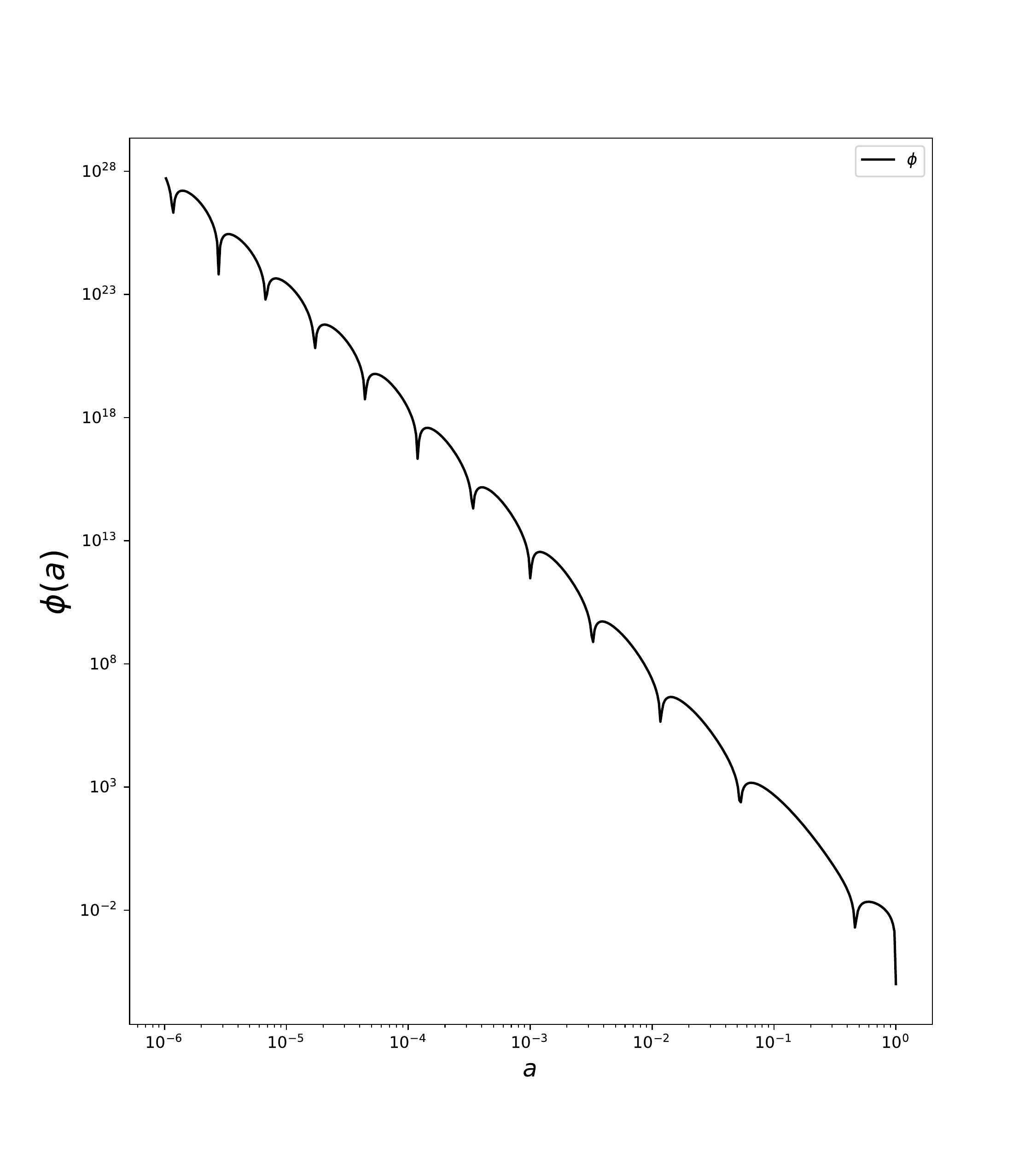}
\includegraphics[width=0.40\textwidth]{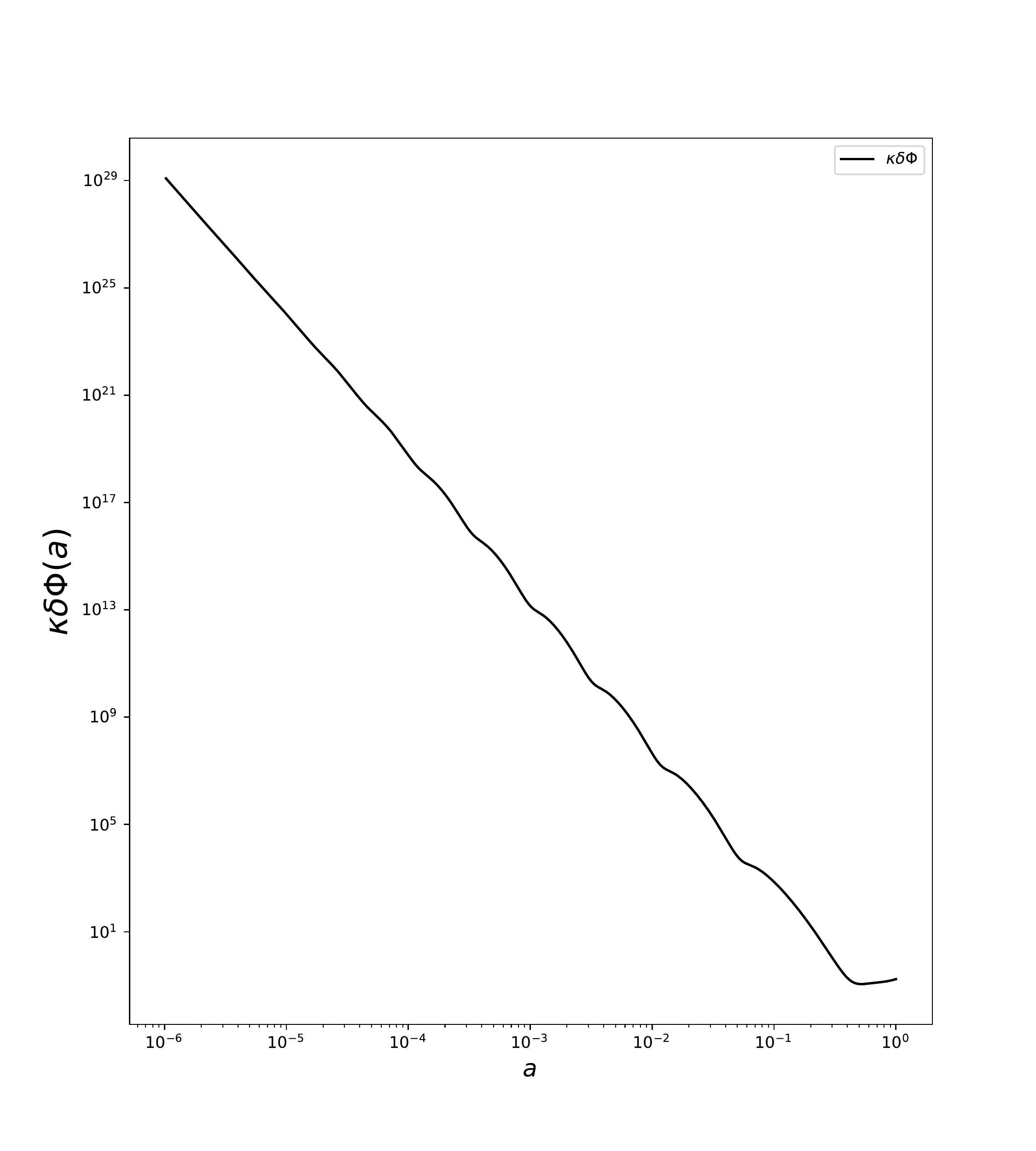}
\caption{The evolution of the absolute value of the Newtonian potential $\phi$ (upper panel) and the evolution of the fluctuation of the scalar field $\delta\Phi$ (lower panel).
Note that in contrast to Fig.\ref{fig:Conqcerophidelphi}, now the Newtonian potential has a strong quantum behavior, while the scalar field perturbation evolves almost without oscillations, exactly the opposite situation as in Fig.\ref{fig:Conqcerophidelphi}. This behavior of $\delta\Phi$ has been shown in several articles in the past, see for example Matos \& Urena-Lopez (2001) and Hlozek et. al. (2014) for a real scalar field and Foidl \& Rindler-Deller (2022) for the complex scalar field. We put it here to compare these results with those presented in  this work.  {Note that the evolution of these functions depends weakly on the character of the scalar field, for real, oscillatons, and for complex scalar fields the evolution is similar. The charge does not play an important role here. This result is also compatible with Tellez-Tovar et. al. (2021)} }
\label{fig:delPhi}
\end{figure}
\begin{figure}
\centering
\includegraphics[width=0.40\textwidth]{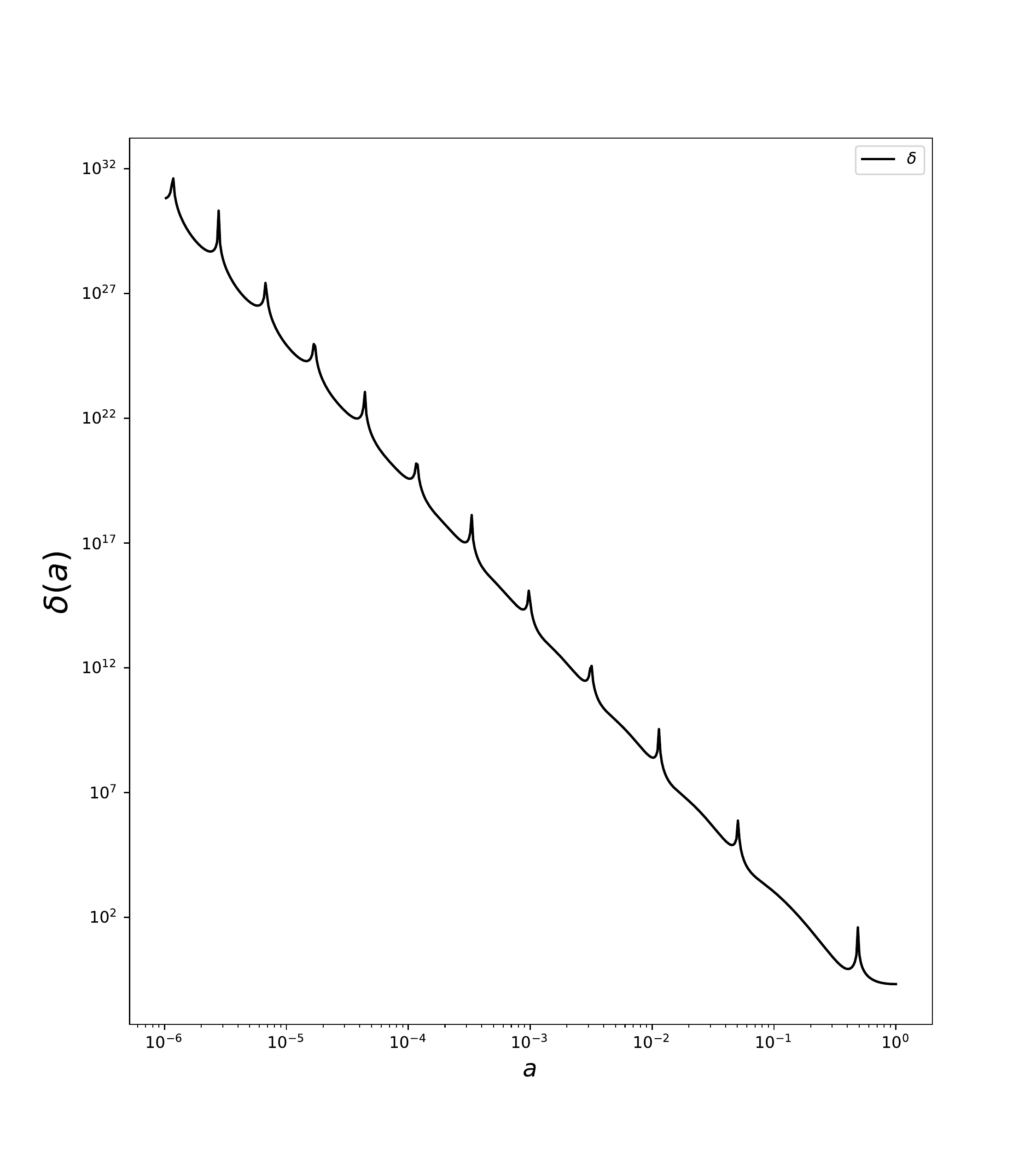}
\includegraphics[width=0.40\textwidth]{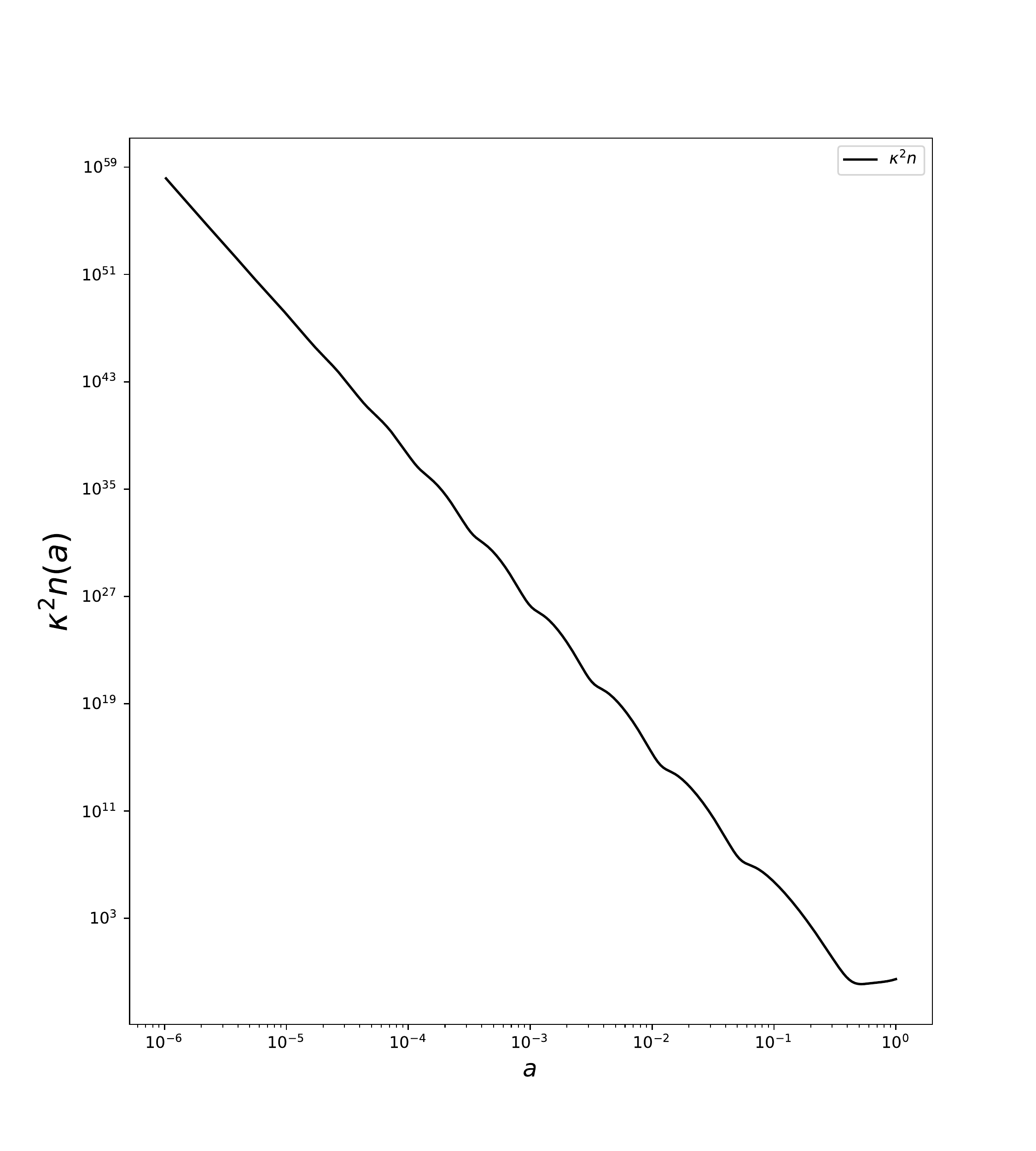}
\caption{The behavior of the absolute value of the density contrast $\delta$ (upper panel) and the numerical density $n$ (lower panel) for the fluctuation of the scalar field. We need to take the absolute value of the density contrast, because it has strong swings from positive to negative due to its quantum nature. Notice that the density contrast oscillates strongly while the number density $n$ hardly oscillates. This is the reason why we can compare the number density with the evolution of the hydrodynamic counterparts, $n$ behaves almost classically.  }
\label{fig:n1}
\end{figure}
\begin{figure}
\centering
\includegraphics[width=0.43\textwidth]{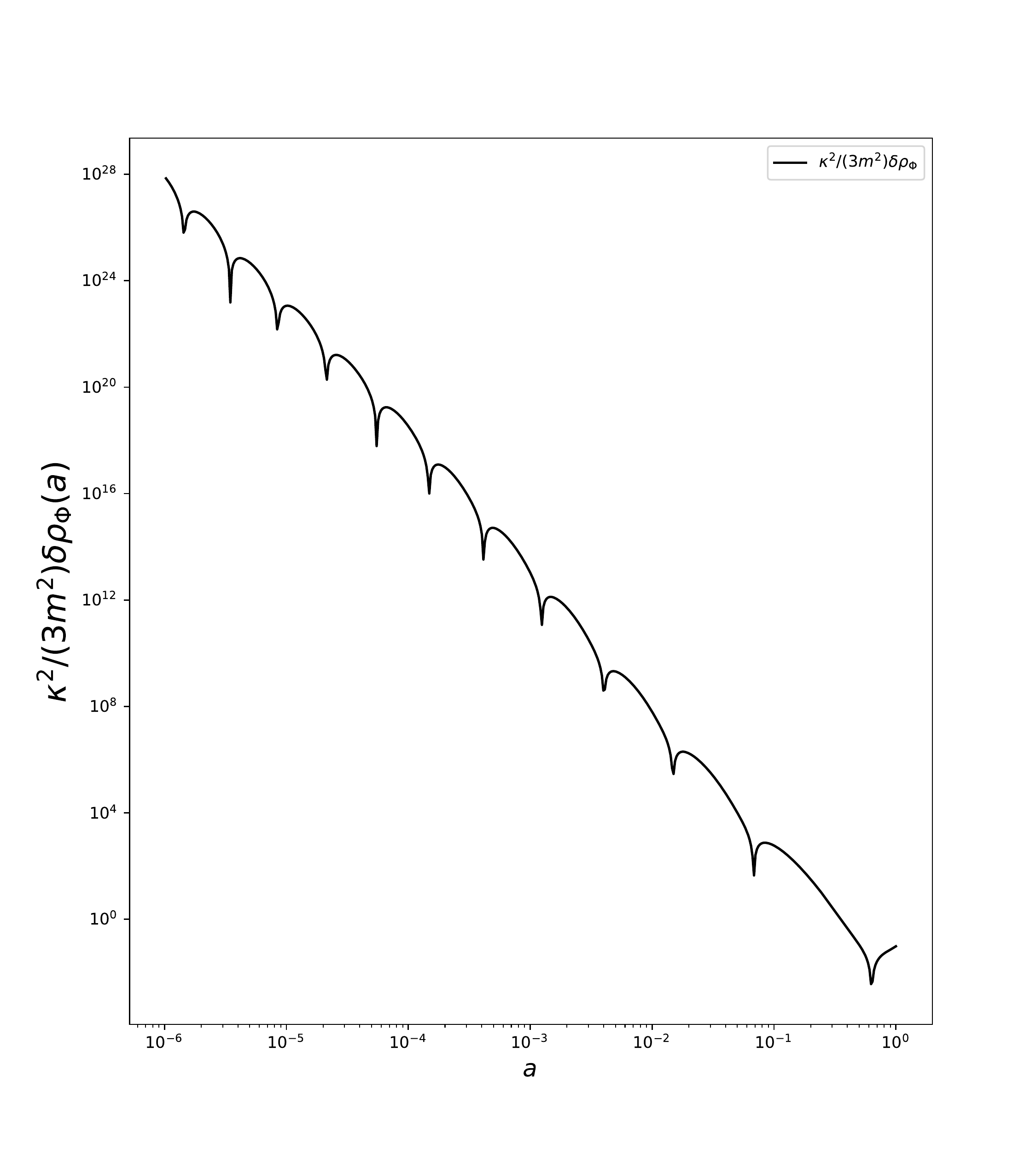}
\caption{In this figure we show the absolute values of the evolution of the fluctuation of the density of the scalar field. Again, the oscillations are a consequence of the influence of the quantum mechanical character of the SFDM. What we can measure is the average of fluctuation of the scalar field, which can be seen as almost classical. }
\label{fig:deltarhoPhi1}
\end{figure}

 In this case, again, the entire system can be solved numerically as usual, taking the Fourier transform and solving the transformed set of differential equations as a dynamical system. But to elucidate the physical behavior of the system, as in the previous example, it is not necessary to solve the system, we will make a toy model to see its behavior, but now with an electromagnetic field. Galaxies generally contain a complicated magnetic field, it has been speculated in \cite{Hernandez:2018cir} that this field could be a consequence of SFDM with the Lagrangian (\ref{eq:Lagrangian}), so the solution of Maxwell's equations could be very complicated for a real magnetic field of a scalar field fluctuation. In addition, the magnetic field does not have a determining influence on the gravitational field of a fluctuation, it is generally very weak in the galaxy. The important fact here is that the Lagrangian (\ref{eq:Lagrangian}) predicts its existence. Therefore, for this toy model we will ignore that the electromagnetic field must be a complicated solution of Maxwell's equations (\ref{eq:Maxwell-d}) and write just a toy electromagnetic field. To solve the toy system, here we will assume that the electromagnetic potential $B_\mu$ can be derived from a superpotential, for this we define two functions, $A$ and $B$, such that $B_{j}=a^2(\dot B\phi)_{,j} $ and $\dot A=\sqrt{n_0} (a^2\sqrt{n_0}B_0\dot)$. These two functions cannot always be defined like this, but the existence of the superpotential for $B_\mu$ can allow them. In this section we limit ourselves to this case. With these definitions, the equations (\ref{eq:ecuacionesdeeinstein2-d}) and (\ref{eq:KGIm-d}) reduce to
 \begin{eqnarray}
      2(\dot{\phi}+H\phi)_{,j}&=&\kappa^2\left(\dot n_0\delta_{,j}+2q\,s_0(\dot B\phi)_{,j}\right),\label{eq:ecuacionesdeeinstein2-dB}\\
       \dot\delta-2\dot\phi+\frac{1}{2s_0}\dot A&=&0.
 \end{eqnarray}

In this case it is possible to integrate (\ref{eq:ecuacionesdeeinstein2-dB}), we get that
\begin{eqnarray}
    \phi=\frac{P_0}{a}e^{\kappa^2(n_0+qs_0B+T(t))},\\
    \delta=2\phi+\frac{1}{2s_0}A+Y(x^i),
\end{eqnarray}
where the function $P_0$ again satisfies the equation (\ref{eq:P0}) with solution (\ref{eq:solP0}) for $l=0$ and the corresponding solutions for $l\not=0$, again similar to an atom, with $Y$ and $T$ arbitrary integration functions. Note that the entire integration of the system depends on the functions $n_0$ and $a$, and of course on how we choose the arbitrary functions $Y$ and $T$. The solutions are in general similar to the previous ones with $B_\mu=0$, if the arbitrary functions are zero. 

\section{The general case}\label{sec:8General}

In the general case we have to add the equations (\ref{eq:ecuacionesdeeinstein1}) and (\ref{eq:KGperturbada}). Here we remark that the electromagnetic field is not essential for the development of the fluctuations, it is essential in the final fluctuation to explain the magnetic field of the galaxies (\cite{Hernandez:2018cir}) and the FB, and for the purpose of this work we can also neglect the proper interactions $\lambda$. So, let's substitute $B_\mu=0$ and take the scalar field potential as $V=m^2\Phi\Phi^\ast$. 

As in the toy models, we separate the function $\phi=P_0T$, where again the function $P_0=P_0(x^i)$, depends only on the spatial coordinates, and $T=T(\eta)$. We make an analogous separation with the fluctuation of the scalar field $\delta\Phi=P_1T_1$. It is easy to see that the equation (\ref{eq:ecuacionesdeeinstein1}) can again be written as
\begin{equation}\label{eq:P01}
     \nabla^2P_0+\Omega^2P_0=K,
\end{equation}
where now
\begin{equation}
     \Omega^2=\kappa^2\dot\Phi_0\dot\Phi_0^*T-3H\left(\dot T+HT\right)
\end{equation}
and $K$ is a small function given by
\begin{equation*}
K=\frac{\kappa^2}{2}\left[\left(\dot\Phi_0 \frac{\dot T_1}{T_1}+a^2m_\Phi^2\Phi_0\right)\delta\Phi ^*+c.c.\right].
\end{equation*}
If the function $K$ remains small, the solutions of the differential equation (\ref{eq:P01}) should be very similar to the previous one representing atoms. Therefore, our conclusion is that, in general, it will be very common for the collapse of the SFDM to be a halo very similar to an atom.

In Fourier space, the equations (\ref{eq:ecuacionesdeeinstein1}) and (\ref{eq:KGperturbada}) read
\begin{eqnarray}
    &&-2k^2\phi_k-6H(\dot{\phi_k}+H\phi_k)=\nonumber\\
&&\kappa^2[(\dot{\Phi}_0\delta\dot\Phi_k^{\ast}+\dot{\Phi}^{\ast}_0\dot{\delta\Phi_k})-2\phi_k\dot{\Phi}_0\dot{\Phi}^{\ast}_0\nonumber\\
&&+a^2m^2(\Phi_0^\ast\delta\Phi_k+\Phi_0\delta\Phi_k^\ast],\label{eq:G00-F}\\
&&-k^2\delta\Phi_k-\ddot{\delta\Phi_k}-2H\dot{\delta\Phi_k}-\frac{iq}{a^2}(a^2\Phi_0B_{0k}\dot)\nonumber\\
&&+4\dot\Phi_0\dot{\phi_k}+m^2a^2\delta\Phi_k-2m^2\Phi_0a^2\phi_k=0.\label{eq:KG-F}
\end{eqnarray}
We now define the unitless variables for the perturbation
\begin{eqnarray}\label{eq:dxdydudv}
    \delta x_k&=&\frac{\kappa}{\sqrt{3}}\frac{\delta\dot\Phi_k}{a\mathcal{H}},\,\,\,\,\,\,\,\,\,\,\,\,\,\,\,\, \delta y_k=\frac{\kappa}{\sqrt{3}}\frac{\delta\dot\Phi_k^\ast}{a\mathcal{H}},\nonumber\\
    \delta u_k&=&\frac{\kappa}{\sqrt{3}}\frac{m\delta\Phi_k}{\mathcal{H}},\,\,\,\,\,\,\,\,\,\, \delta v_k=\frac{\kappa}{\sqrt{3}}\frac{m\delta\Phi_k^\ast}{\mathcal{H}},
\end{eqnarray}
we find that the equations (\ref{eq:G00-F}) and (\ref{eq:KG-F}) become
\begin{eqnarray}
    2\frac{k^2}{m^2}s^2\phi_k+6({\phi'_k}+\phi_k)&=&\nonumber\\
-3(x\delta y_k+y\delta x_k)+6xy\phi_k&-&3(u\delta v_k+v\delta u_k),\nonumber\\
\left(-\frac{k^2}{m^2}+1\right)s\delta u_k-\delta  x'_k&-&\left(\frac{3}{4}\Pi+1\right)\delta x_k\nonumber\\
+4x\phi'_k-su\phi_k&=&0.
\end{eqnarray}
We can also write the scalar field density (\ref{eq:rhoPhi0}) and the scalar field density fluctuation (\ref{eq:deltarho}) in terms of these variables, we get
\begin{eqnarray}
    \frac{\kappa^2}{3m^2}\rho_{\Phi_0}&=&\frac{1}{s^2}(xy+uv),\\
    \frac{\kappa^2}{3m^2}\delta\rho_{\Phi}&=&-\frac{1}{s^2}\left(x\delta y_k+y\delta x_k-2\phi_k xy+u\delta v_k+v\delta u_k\right).\nonumber\\
\end{eqnarray}

The system (\ref{eq:xpypupvp}), together with the previous equations, can be integrated numerically. The results are given in the figures. In Fig.\ref{fig:delPhi} we see the evolution of the absolute value of the fluctuation of the scalar field $\delta\Phi$. We have to plot the absolute value because the oscillations go from very large positive values to very large negative values. The absolute value is presented only to see the behavior, an observer will measure the average of this behavior. In Fig.\ref{fig:n1} we represent the absolute value of the density contrast $\delta$ and the evolution of the numerical density $n$. Here we see just the opposite of Fig.\ref{fig:Conqcerophidelphi} where the density contrast for the exact solution evolves smoothly, here its quantum character predominates showing strong oscillations. The number density $n$ here oscillates very little, so the evolution is similar to that of a fluid.

Finally in Fig.\ref{fig:deltarhoPhi1} we see the absolute value of the evolution of the density fluctuation of the scalar field $\delta\rho_\Phi$. As also shown in \cite{Magana:2012xe} for a real scalar field, here the density fluctuation of the complex scalar field oscillates strongly. However, an observer will only detect the average of this oscillation, whose behavior resembles that of a fluid. Therefore the SFDM can be confused with that of a fluid in its behavior. This important result is also valid for the charged complex scalar field. 

\section{Conclusions}\label{sec:9conclusions}

In the present work we have established the bases for a complete model of dark matter where the nature of it is an ultralight scalar field, that is, a particle with zero spin and an extremely small mass, with an extremely small self-interaction. So that this scalar field could explain most of the observed phenomena in a galaxy, such as the core centers of galaxies, the small number of satellite galaxies around their hosts, the VPO or polar orbits of satellite galaxies around their host, the magnetic fields of galaxies, the FB, etc., in addition to the excellent fit with cosmological observations, we have to propose a charged scalar field. The real scalar field cannot be charged, so we must start with a complex scalar field whose Lagrangian is invariant under the group $U(1)$. However, there are two possibilities.

1.- The corresponding photon of the complex scalar field is the SM photon. In this case, the charge must be ultralight to be in agreement with the observational constrains in this respect that we have so far.

2.- The corresponding photon interacts with the SM photon but not with the rest of the matter, the so-called Dark Photon.

In this work we have established the field equations for the cosmological evolution of the charged scalar field fluctuations. On cosmological scales we do not see that dark matter can be charged, we do not have any evidence of it, so we conclude that the charge can only be generated as a fluctuation that we consider to be small, due to observations in galaxies. This charge could be the origin of the magnetic field observed in galaxies, as speculated in \cite{Hernandez:2018cir}. In this work, using the ansatz (\ref{eq:deltaPhi2}) and (\ref{eq:deltaPhi3}) we were able to integrate the perturbed field equations in terms of the two background functions $a$ and $n_0$, which can be easily integrated numerically. We did it without and with the electromagnetic field. The interesting point here is that these two toy models show us the fact that the space-time configuration of the final SFDM halo is the same as that of an atom. This point is important, because this result is replicated in realistic SFDM halo formation using numerical simulations. The main conclusions we have, then, is that the SFDM halos of galaxies are real atoms and thus have, like an atom, different excited states that must be taken into account. This result is precisely the reason why the SFDM can explain the VPO and the FB in galaxies. This result, of course, has consequences. The first is that if this result is correct, we have to see the VPO phenomenon in many more galaxies. The second is that FB must also be present in these galaxies and that the energy of the flashes that we see now as X-rays or gamma rays must be even more energetic and we should be able to see them with the right instruments in the near future.

In conclusion, the quantum nature of the scalar field is the fundamental difference between SFDM and other models, and it is precisely this characteristic that can explain some phenomena observed in galaxies that other models are unable to explain. Quantum mechanics was developed to explain the microworld, but here we conclude that quantum mechanics might also be able to explain the dark side of the cosmos.

\section{acknowledgments}
I want to thank Laura Luz Parrilla for helpful discussions and Luis Osvaldo T\'ellez for his advice on python codes. This work was partially supported by CONACyT M\'exico under grants  A1-S-8742, 304001, 376127 and
FORDECYT-PRONACES grant No. 490769\\

Data availability: There are no new data associated with this article.

\bibliographystyle{mnras}
\bibliography{QuantumMNRAS}

\begin{thebibliography}{}

\bibitem{Suarez:2013iw}
A.~Su\'arez, V.~H.~Robles and T.~Matos,
Astrophys. Space Sci. Proc. \textbf{38} (2014), 107-142
doi:10.1007/978-3-319-02063-1\_9
[arXiv:1302.0903 [astro-ph.CO]].

\bibitem{Oks:2021hef}
E.~Oks,
New Astron. Rev. \textbf{93} (2021), 101632
doi:10.1016/j.newar.2021.101632
[arXiv:2111.00363 [astro-ph.CO]].

\bibitem{Matos:1998vk}
T.~Matos and F.~S.~Guzman,
Class. Quant. Grav. \textbf{17} (2000), L9-L16
doi:10.1088/0264-9381/17/1/102
[arXiv:gr-qc/9810028 [gr-qc]].

\bibitem{Hu:2000ke}
W.~Hu, R.~Barkana and A.~Gruzinov,
Phys. Rev. Lett. \textbf{85} (2000), 1158-1161
doi:10.1103/PhysRevLett.85.1158
[arXiv:astro-ph/0003365 [astro-ph]].

\bibitem{Boehmer:2007um}
C.~G.~Boehmer and T.~Harko,
JCAP \textbf{06} (2007), 025
doi:10.1088/1475-7516/2007/06/025
[arXiv:0705.4158 [astro-ph]].

\bibitem{Rindler-Daller:2009qyu}
T.~Rindler-Daller and P.~R.~Shapiro,
ASP Conf. Ser. \textbf{432} (2010), 244
[arXiv:0912.2897 [astro-ph.CO]].

\bibitem{Bray:2010fc}
H.~L.~Bray,
[arXiv:1004.4016 [physics.gen-ph]].

\bibitem{Schive:2014dra}
H.~Y.~Schive, T.~Chiueh and T.~Broadhurst,
Nature Phys. \textbf{10} (2014), 496-499
doi:10.1038/nphys2996
[arXiv:1406.6586 [astro-ph.GA]].

\bibitem{Hui:2016ltb}
L.~Hui, J.~P.~Ostriker, S.~Tremaine and E.~Witten,
Phys. Rev. D \textbf{95} (2017) no.4, 043541
doi:10.1103/PhysRevD.95.043541
[arXiv:1610.08297 [astro-ph.CO]].

\bibitem{Matos:2000ss}
T.~Matos and L.~A.~Urena-Lopez,
Phys. Rev. D \textbf{63} (2001), 063506
doi:10.1103/PhysRevD.63.063506
[arXiv:astro-ph/0006024 [astro-ph]].

\bibitem{Hlozek:2014lca}
R.~Hlozek, D.~Grin, D.~J.~E.~Marsh and P.~G.~Ferreira,
Phys. Rev. D \textbf{91} (2015) no.10, 103512
doi:10.1103/PhysRevD.91.103512
[arXiv:1410.2896 [astro-ph.CO]].

\bibitem{Guzman:2003kt}
F.~S.~Guzman and L.~A.~Urena-Lopez,
Phys. Rev. D \textbf{68} (2003), 024023
doi:10.1103/PhysRevD.68.024023
[arXiv:astro-ph/0303440 [astro-ph]].

\bibitem{Urena-Lopez:2002nup}
L.~A.~Urena-Lopez and A.~R.~Liddle,
Phys. Rev. D \textbf{66} (2002), 083005
doi:10.1103/PhysRevD.66.083005
[arXiv:astro-ph/0207493 [astro-ph]].

\bibitem{Avilez:2017jql}
A.~A.~Avilez, L.~E.~Padilla, T.~Bernal-Marin and T.~Matos,
Mon. Not. Roy. Astron. Soc. \textbf{477} (2018) no.3, 3257-3272
doi:10.1093/mnras/sty572
[arXiv:1704.07314 [gr-qc]].

\bibitem{Padilla:2020sjy}
L.~E.~Padilla, T.~Rindler-Daller, P.~R.~Shapiro, T.~Matos and J.~Alberto V\'azquez,
Phys. Rev. D \textbf{103} (2021) no.6, 063012
doi:10.1103/PhysRevD.103.063012
[arXiv:2010.12716 [astro-ph.GA]].

\bibitem{Barranco:2011eyw}
J.~Barranco, A.~Bernal, J.~C.~Degollado, A.~Diez-Tejedor, M.~Megevand, M.~Alcubierre, D.~Nunez and O.~Sarbach,
Phys. Rev. D \textbf{84} (2011), 083008
doi:10.1103/PhysRevD.84.083008
[arXiv:1108.0931 [gr-qc]].

\bibitem{Alcubierre:2018ahf}
M.~Alcubierre, J.~Barranco, A.~Bernal, J.~C.~Degollado, A.~Diez-Tejedor, M.~Megevand, D.~Nunez and O.~Sarbach,
Class. Quant. Grav. \textbf{35} (2018) no.19, 19LT01
doi:10.1088/1361-6382/aadcb6
[arXiv:1805.11488 [gr-qc]].

\bibitem{Urena-Lopez:2010zva}
L.~A.~Urena-Lopez and A.~Bernal,
Phys. Rev. D \textbf{82} (2010), 123535
doi:10.1103/PhysRevD.82.123535
[arXiv:1008.1231 [gr-qc]].

\bibitem{Guzman:2019gqc}
F.~S.~Guzm\'an and L.~A.~Ure\~na-L\'opez,
Phys. Rev. D \textbf{101} (2020) no.8, 081302
doi:10.1103/PhysRevD.101.081302
[arXiv:1912.10585 [astro-ph.GA]].

\bibitem{Guzman:2022vxl}
F.~S.~Guzman,
Phys. Rev. D \textbf{105} (2022) no.12, 123535
doi:10.1103/PhysRevD.105.123535
[arXiv:2206.03407 [gr-qc]].

\bibitem{Matos:2011pd}
T.~Matos and A.~Su\'arez,
Class. Quant. Grav. \textbf{31} (2014) no.4, 045015
doi:10.1088/0264-9381/31/4/045015
[arXiv:1103.5731 [gr-qc]].

\bibitem{Robles:2012kt}
V.~H.~Robles and T.~Matos,
Astrophys. J. \textbf{763} (2013), 19
doi:10.1088/0004-637X/763/1/19
[arXiv:1207.5858 [astro-ph.CO]].

\bibitem{Robles:2013ioa}
V.~H.~Robles and T.~Matos,
Springer Proc. Phys. \textbf{148} (2013), 17-24
doi:10.1007/978-94-007-7241-0\_2

\bibitem{Rindler-Daller:2013zxa}
T.~Rindler-Daller and P.~R.~Shapiro,
Mod. Phys. Lett. A \textbf{29} (2014) no.2, 1430002
doi:10.1142/S021773231430002X
[arXiv:1312.1734 [astro-ph.CO]].

\bibitem{Pawlowski:2013cae}
M.~S.~Pawlowski and P.~Kroupa,
Mon. Not. Roy. Astron. Soc. \textbf{435} (2013), 2116
doi:10.1093/mnras/stt1429
[arXiv:1309.1159 [astro-ph.CO]].

\bibitem{Pawlowski:2019bar}
M.~S.~Pawlowski and P.~Kroupa,
Mon. Not. Roy. Astron. Soc. \textbf{491} (2020) no.2, 3042-3059
doi:10.1093/mnras/stz3163
[arXiv:1911.05081 [astro-ph.GA]].

\bibitem{Ibata:2013rh}
R.~A.~Ibata, G.~F.~Lewis, A.~R.~Conn, M.~J.~Irwin, A.~W.~McConnachie, S.~C.~Chapman, M.~L.~Collins, M.~Fardal, A.~M.~N.~Ferguson and N.~G.~Ibata, \textit{et al.}
Nature \textbf{493} (2013), 62-65
doi:10.1038/nature11717
[arXiv:1301.0446 [astro-ph.CO]].

\bibitem{Conn:2013iu}
A.~R.~Conn, G.~F.~Lewis, R.~A.~Ibata, Q.~A.~Parker, D.~B.~Zucker, A.~W.~McConnachie, N.~F.~Martin, D.~Valls-Gabaud, N.~Tanvir and M.~J.~Irwin, \textit{et al.}
Astrophys. J. \textbf{766} (2013), 120
doi:10.1088/0004-637X/766/2/120
[arXiv:1301.7131 [astro-ph.CO]].

\bibitem{Muller:2018hks}
O.~M\"uller, M.~S.~Pawlowski, H.~Jerjen and F.~Lelli,
Science \textbf{359} (2018), 534
doi:10.1126/science.aao1858
[arXiv:1802.00081 [astro-ph.GA]].

\bibitem{Muller:2020njt}
O.~M\"uller, M.~S.~Pawlowski, F.~Lelli, K.~Fahrion, M.~Rejkuba, M.~Hilker, J.~Kanehisa, N.~Libeskind and H.~Jerjen,
Astron. Astrophys. \textbf{645} (2021), L5
doi:10.1051/0004-6361/202039973
[arXiv:2012.08138 [astro-ph.GA]].

\bibitem{Su:2010qj}
M.~Su, T.~R.~Slatyer and D.~P.~Finkbeiner,
Astrophys. J. \textbf{724} (2010), 1044-1082
doi:10.1088/0004-637X/724/2/1044
[arXiv:1005.5480 [astro-ph.HE]].

\bibitem{Li:2013nal}
B.~Li, T.~Rindler-Daller and P.~R.~Shapiro,
Phys. Rev. D \textbf{89} (2014) no.8, 083536
doi:10.1103/PhysRevD.89.083536
[arXiv:1310.6061 [astro-ph.CO]].

\bibitem{Solis-Lopez:2019lvz}
J.~Sol\'\i{}s-L\'opez, F.~S.~Guzm\'an, T.~Matos, V.~H.~Robles and L.~A.~Ure\~na-L\'opez,
Phys. Rev. D \textbf{103} (2021) no.8, 083535
doi:10.1103/PhysRevD.103.083535
[arXiv:1912.09660 [astro-ph.GA]].

\bibitem{DES:2020fxi}
E.~O.~Nadler \textit{et al.} [DES],
Phys. Rev. Lett. \textbf{126} (2021), 091101
doi:10.1103/PhysRevLett.126.091101
[arXiv:2008.00022 [astro-ph.CO]].

\bibitem{Matos:2022ruq}
T.~Matos, A.~Perez-Lorenzana and J.~Sol\'\i{}s-L\'opez,
[arXiv:2203.13218 [astro-ph.GA]].

\bibitem{Urena-Lopez:2015gur}
L.~A.~Ure\~na-L\'opez and A.~X.~Gonzalez-Morales,
JCAP \textbf{07} (2016), 048
doi:10.1088/1475-7516/2016/07/048
[arXiv:1511.08195 [astro-ph.CO]].

\bibitem{Medellin-Gonzalez:2020eww}
S.~G.~Medell\'\i{}n-Gonz\'alez, L.~Arturo Ure\~na-L\'opez and A.~X.~Gonz\'alez-Morales,
Phys. Rev. D \textbf{103} (2021) no.8, 083509
doi:10.1103/PhysRevD.103.083509
[arXiv:2010.13998 [astro-ph.CO]].

\bibitem{Foidl:2022bpn}
H.~Foidl and T.~Rindler-Daller,
Phys. Rev. D \textbf{105} (2022) no.12, 123534
doi:10.1103/PhysRevD.105.123534
[arXiv:2203.09396 [astro-ph.CO]].

\bibitem{Matos:2016ryp}
T.~Matos, A.~Avilez, T.~Bernal and P.~H.~Chavanis,
Gen. Rel. Grav. \textbf{51} (2019) no.12, 159
doi:10.1007/s10714-019-2644-9
[arXiv:1608.03945 [gr-qc]].

\bibitem{Magana:2012xe}
J.~Magana, T.~Matos, A.~Suarez and F.~J.~Sanchez-Salcedo,
JCAP \textbf{10} (2012), 003
doi:10.1088/1475-7516/2012/10/003
[arXiv:1204.5255 [astro-ph.CO]].

\bibitem{Hernandez:2018cir}
M.~Hernandez, A.~Avilez-L\'opez and T.~Matos,
Eur. Phys. J. C \textbf{79} (2019) no.10, 833
doi:10.1140/epjc/s10052-019-7349-4
[arXiv:1805.09930 [gr-qc]].

\bibitem{Bohm:1951xx}
D.~Bohm,
Phys. Rev. \textbf{85} (1952), 180-193
doi:10.1103/PhysRev.85.180

\bibitem{Tellez-Tovar:2021mge}
L.~O.~T\'ellez-Tovar, T.~Matos and J.~A.~V\'azquez,
[arXiv:2112.09337 [astro-ph.CO]].

\end{thebibliography}

\section{Appendix A. The field equations in terms of the unitless variables} 

Using definitions (\ref{eq:defxyuv}) we can extract the different values of the physical variables. First observe that we have
\begin{equation}\label{eq:n0uvs}
 uv=\frac{\kappa^2}{3}\frac{m^2}{\mathcal{H}^2}n_0=\frac{\kappa^2}{3}s^2n_0;\,\,\,\,\,\,\,\,\,\,\frac{u}{v}=e^{2i\theta_0
 }   
\end{equation}
and $\Phi_0=\sqrt{3}u/s$.

If $B_\mu=0$, using (\ref{eq:solP0}) as $\delta=2\phi$ and $n= n_0\delta$ we have
\begin{equation}
 \phi=P_0\exp(\frac{3uv}{s^2}-N),\,\,\,\,\,\,\,\,\,\,n=\frac{2P_0}{\kappa^2}\frac{3uv}{s^2}\exp(\frac{3uv}{s^2}-N).
\end{equation}
On the other hand, $\delta\Phi=\Phi_0\delta=\sqrt{n}e^{i\theta_0}$, thus
\begin{equation}
    \delta\Phi=\frac{\sqrt{6P_0}}{\kappa}\frac{u}{s}\exp(\frac{3uv}{2s^2}-\frac{1}{2}N).
\end{equation}

Finally we calculate the quantum potential, in (\ref{eq:UQ}) the quantum potential is defined as
\begin{equation*}
    U_0^Q=-\frac{1}{2m^2}\frac{\Box\sqrt{n_0}}{\sqrt{n_0}}.
\end{equation*}
In this work we are interested in the quantum evolution of the background, therefore the D'Alambertian reduce to the calculation of the time evolution $\Box\sqrt{n_0}=-\ddot{\sqrt{n_0}}$. In order to see this value, we use the expression for $n_0$ in (\ref{eq:n0uvs}) and then derive it with respect to $N$. Using equations (\ref{eq:xpypupvp}) we obtain that
\begin{equation}
    U_0^Q=-\frac{3m^2}{\kappa^2\,s^3}(3(uy+vx)+2s(vu-xy)).
\end{equation}

In the general case, using variables (\ref{eq:dxdydudv}) and that $ \kappa^2 n_0={3uv}/{s^2}$, we have
\begin{eqnarray}
   \delta_k&=&\frac{\delta u_k}{u},\,\,\,\,\,\,\,\,\,\,\delta\Phi_k=\sqrt{3}\frac{\delta u_k}{s},\nonumber\\
   \kappa^2 n_k&=&3\frac{\delta u_k^2}{u}\frac{v}{s^2}.
\end{eqnarray}

We take the scalar field potential as $V=m^2\Phi\Phi^\ast$. Then, equations (\ref{eq:KG-dk}) and (\ref{eq:phik}) reduce to
\begin{eqnarray}
    &&\ddot\delta_k+2\left(H+\frac{\dot\Phi_0}{\Phi_0}\right)\dot\delta_k +k^2\delta_k-4\left({\dot\phi}_k+H\phi_k\right)\frac{\dot\Phi_0}{\Phi_0}\nonumber\\
    &&+2\left(2H\frac{\dot\Phi_0}{\Phi_0}+m^2a^2\right)\phi_k=0,\\
    &&\ddot{\phi_k}+6H\dot{\phi_k}+2\phi_k\left(\dot{H}+2H^2\right)+\frac{k^2}{a^2}\phi_k\nonumber\\
    &&+2\kappa^2\,m^2\Phi_0\Phi_0^{\ast}\delta_k =0.
\end{eqnarray}
We can rewritte the previous equations in the $N$ variable, taking into account that $\ddot x=H^2(x''-3/2\Pi x')$. We obtain
\begin{eqnarray}\label{eq:KG-dksE}
    &&\delta''_k+2\left(1+\frac{sx}{u}-\frac{3}{2}\Pi \right)\delta'_k +\frac{k^2}{m^2}s^2\delta_k\nonumber\\
    &&-4\left({\phi'}_k+\phi_k\right)\frac{sx}{u}+2\left(2\frac{sx}{u}+s^2e^{2N}\right)\phi_k=0,\\
    &&\phi''_k +3(2-\frac{1}{2}\Pi)\phi'_k+2\left(2-\frac{3}{2}\Pi\right)\phi_k+\frac{k^2}{m^2}s^2e^{-2N}\phi_k\nonumber\\
    &&+6uv\delta_k =0.
\end{eqnarray}
%











\end{document}